\documentclass[acmlarge]{acmart}

\setcopyright{none}
\renewcommand\footnotetextcopyrightpermission[1]{}

\AtBeginDocument{%
  }

\setcopyright{acmcopyright}
\copyrightyear{2025}
\acmYear{2025}
\acmDOI{1111111.1111111}

\acmJournal{IMWUT}
\acmVolume{0}
\acmNumber{0}
\acmArticle{000}
\acmMonth{0}

\usepackage[english]{babel}
\usepackage{booktabs} 
\usepackage[absolute,overlay]{textpos}
\usepackage{etoolbox}
\usepackage[figurename=Fig.,font={small},labelfont={up}]{caption}
\usepackage[export]{adjustbox}
\usepackage[linesnumbered,ruled,vlined]{algorithm2e}
\usepackage{color, colortbl}
\usepackage{soul}
\usepackage{comment}
\usepackage{url}
\usepackage{balance}
\usepackage{graphicx}
\usepackage{subfigure}
\usepackage{multirow}
\usepackage{textcomp}
\usepackage{lipsum}
\usepackage{tikz}
\usepackage{hyperref}
\usepackage[capitalise]{cleveref}
\usepackage{float}
\usepackage{xspace}
\usepackage{enumitem}
\usepackage{tabularx}
\usepackage{lscape}
\usepackage{tcolorbox}
\usepackage{makecell}

\newcommand{\etal}{\textit{et al.}}
\newcommand{\eg}{\textit{e.g.}}
\newcommand{\ie}{\textit{i.e.}}
\usepackage{diagbox}

\newbool{showComments}
\booltrue{showComments}

\ifbool{showComments}{%

}

\begin{document}
\title{A Survey of Earable Technology: Trends, Tools, and the Road Ahead}

\ccsdesc[500]{Human-centered computing~Ubiquitous and mobile computing systems and tools}
\keywords{Earables, Sensing, Health Monitoring, Interactions, Activity, User Authentication}

\author{Changshuo Hu}
\email{cs.hu.2023@phdcs.smu.edu.sg}
\orcid{0000-0002-9432-6073}
\authornote{First authors with equal contribution in alphabetical order.}
\affiliation{%
  \institution{Singapore Management University}
  \country{Singapore}
}

\author{Qiang Yang}
\email{qy258@cam.ac.uk}
\orcid{0000-0002-5202-7892}
\authornotemark[1]
\affiliation{%
  \institution{University of Cambridge}
  \country{United Kingdom}
}

\author{Yang Liu}
\email{yl868@cam.ac.uk}
\orcid{0000-0002-2474-2004}
\authornotemark[1]
\affiliation{%
  \institution{University of Cambridge}
  \country{United Kingdom}
}

\author{Tobias Röddiger}
\email{tobias.roeddiger@kit.edu}
\orcid{0000-0002-4718-9280}
\affiliation{%
  \institution{Karlsruhe Institute of Technology}
  \country{Germany}
}

\author{Kayla-Jade Butkow}
\email{kjb85@cam.ac.uk}
\orcid{0000-0002-5508-7188}
\affiliation{%
  \institution{University of Cambridge}
  \country{United Kingdom}
}

\author{Mathias Ciliberto}
\email{mc2514@cam.ac.uk}
\orcid{0000-0001-9550-7637}
\affiliation{%
  \institution{University of Cambridge}
  \country{United Kingdom}
}

\author{Adam Luke Pullin}
\email{alp@cam.ac.uk}
\orcid{}
\affiliation{%
  \institution{University of Cambridge}
  \country{United Kingdom}
}

\author{Jake Stuchbury-Wass}
\email{js2372@cam.ac.uk}
\orcid{0000-0001-6733-0504}
\affiliation{%
  \institution{University of Cambridge}
  \country{United Kingdom}
}

\author{Mahbub Hassan}
\email{mahbub.hassan@unsw.edu.au }
\orcid{0000-0002-3417-8590}
\affiliation{%
  \institution{University of New South Wales}
  \country{Australia}
}

\author{Cecilia Mascolo}
\email{cm542@cam.ac.uk}
\orcid{0000-0001-9614-4380}
\affiliation{%
  \institution{University of Cambridge}
  \country{United Kingdom}
}

\author{Dong Ma}
\email{dongma@smu.edu.sg}
\orcid{0000-0003-3824-234X}
\affiliation{%
  \institution{Singapore Management University}
  \country{Singapore}
}
\authornote{Corresponding author}
\renewcommand{\shortauthors}{Hu et al.}

\begin{abstract}
Earable devices, wearables positioned in or around the ear, are undergoing a rapid transformation from audio-centric accessories into multifunctional systems for interaction, contextual awareness, and health monitoring. This evolution is driven by commercial trends emphasizing sensor integration and by a surge of academic interest exploring novel sensing capabilities. Building on the foundation established by earlier surveys, this work presents a timely and comprehensive review of earable research published since 2022. We analyze over one hundred recent studies to characterize this shifting research landscape, identify emerging applications and sensing modalities, and assess progress relative to prior efforts. In doing so, we address three core questions: how has earable research evolved in recent years, what enabling resources are now available, and what opportunities remain for future exploration. Through this survey, we aim to provide both a retrospective and forward-looking view of earable technology as a rapidly expanding frontier in ubiquitous computing. In particular, this review reveals that over the past three years, researchers have discovered a variety of novel sensing principles, developed many new earable sensing applications, enhanced the accuracy of existing sensing tasks, and created substantial new resources to advance research in the field. Based on this, we further discuss open challenges and propose future directions for the next phase of earable research.
\end{abstract}
\maketitle

\section{Introduction}

Earables are wearable computing devices worn in or around the ear, capable of supporting audio playback, user interaction, and physiological sensing. Among various earable form factors, true wireless stereo (TWS) earbuds have become the most commercially successful and widely adopted platform~\cite{roddiger2022sensing}. The global market for TWS earbuds has grown exponentially and is expected to continue accelerating in the coming years. Forecasts from The Business Research Company project that the market will expand from 89.6 billion USD in 2024 to 121.91 billion USD in 2025, and further reach 415.69 billion USD by 2029, with a compound annual growth rate of 35.9\% \cite{EarbudsMarket}. This rapid growth is driven not only by the widespread adoption of earphones but also by ongoing technological advances from the academic community, the integration of diverse sensors, and a growing awareness of health and fitness~\cite{butkow2023heart,liu2024respear}.
These trends reflect a fundamental shift: \textit{earbuds are evolving from basic audio playback devices into multifunctional, intelligent systems that support entertainment, communication, and personal wellness}.

\begin{figure}[t]
  \centering

  \begin{minipage}[b]{0.48\textwidth}
    \centering
    \includegraphics[width=\textwidth]{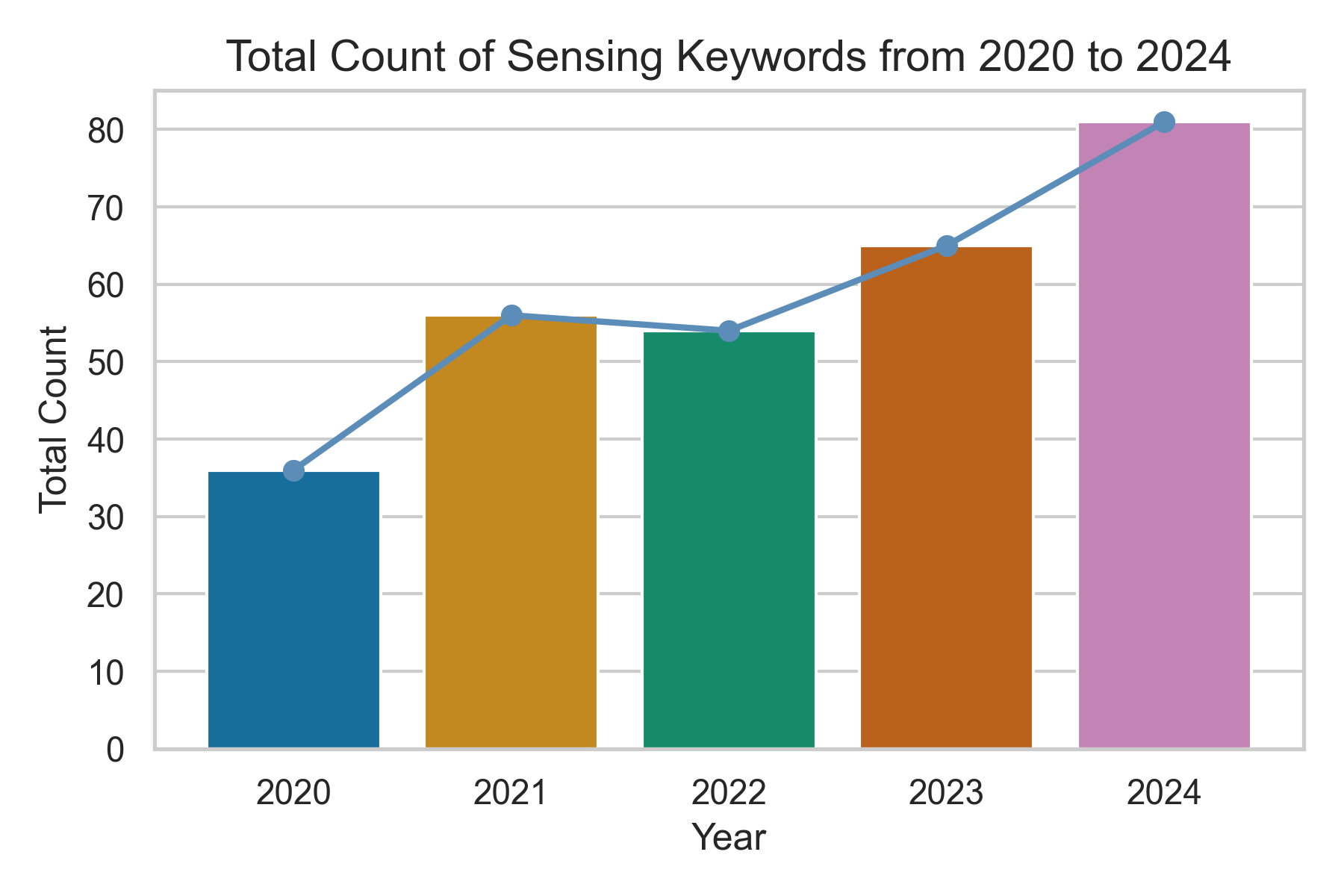}
    \caption{Annual trend of sensing-related keywords mentioned in earable product descriptions from 2020 to 2024.
    } 
    \label{fig: keywordTrend}
  \end{minipage}
  \hfill
  \begin{minipage}[b]{0.48\textwidth}
    \centering
    \includegraphics[width=\textwidth]{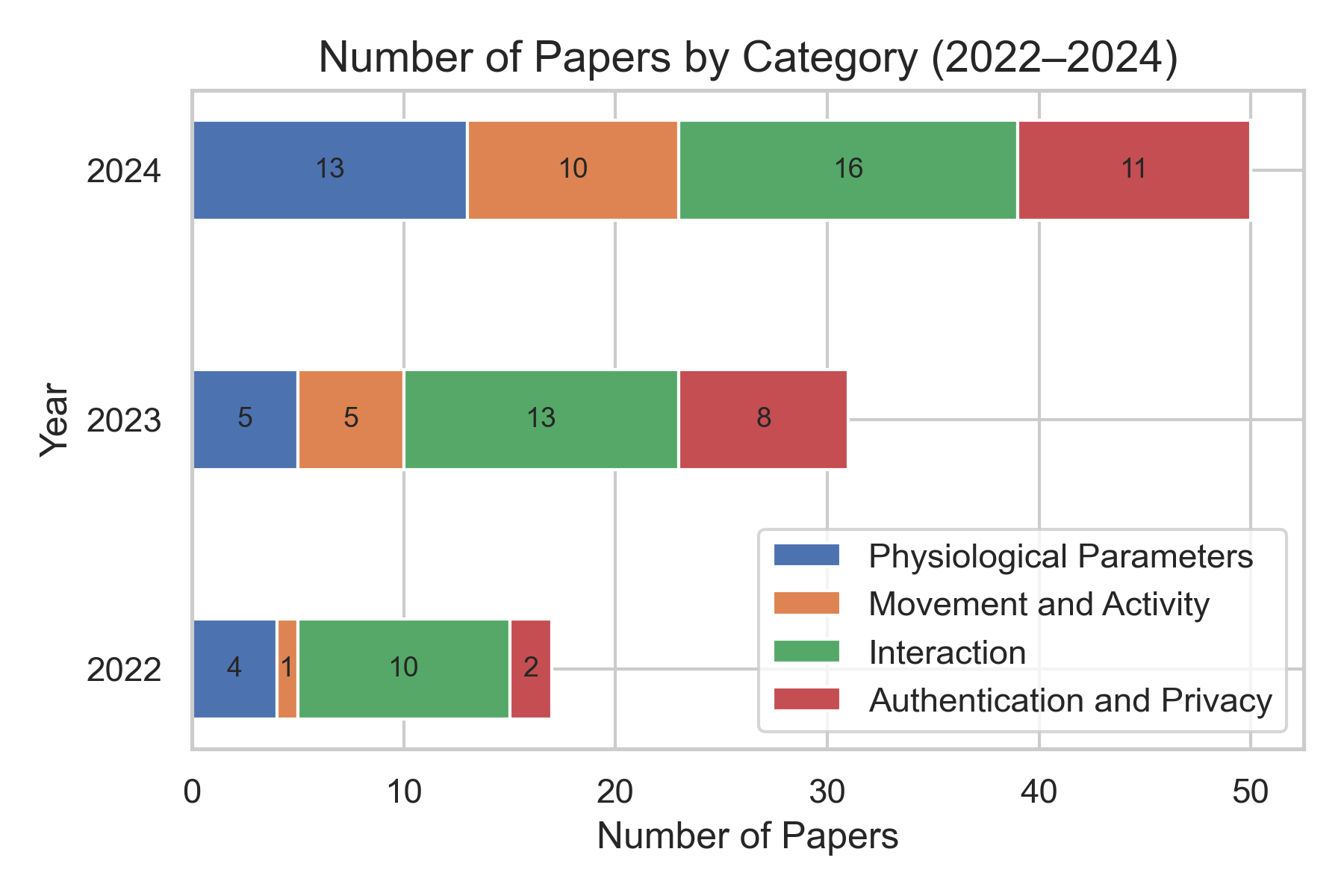}
    \caption{Total number of papers by categories published on earable technologies from 2022 to 2024. 
    }
    \label{fig: paperTrend}
  \end{minipage}
\end{figure}

To better understand how commercial priorities for earbuds have evolved, we analyzed promotional materials for 236 TWS models
released between 2020 and 2025 by major electronics companies (e.g., Apple, Samsung, Sony, Huawei) as well as specialized audio brands (e.g., Bose, Beats, JBL, Sennheiser). For each earbud model, we extracted 4–8 keywords from the promotional content with a focus on product features.
We then identified and counted the number of keywords related to contextual and human-centric sensing\footnote{These sensing-related keywords include: Adaptive Sound Adjustment, Environmental Awareness, Fitness Tracking, Hearing Health, Heart Rate Monitoring, Sleep Monitoring, SpO$_2$ Detection, Spatial Audio, Speech Enhancement, Stress Monitoring, Temperature Monitoring, Touch Control, and Voice Assistant.}.
\Cref{fig: keywordTrend} illustrates the annual trend in the prevalence of these sensing-related keywords. We observe a clear upward trajectory, indicating that manufacturers are increasingly emphasizing sensing and context-awareness features as key differentiators in the TWS market. This shift highlights a broader transition from audio-only devices toward multi-functional, sensor-rich wearables that support diverse applications in health, fitness, and ambient intelligence.

This transformation is grounded in several intrinsic advantages that make earbuds an ideal platform for both user interaction, behavioral sensing, and physiological monitoring. First, as devices primarily used for audio playback and telephony, earbuds are already familiar to users, who naturally interact with them through voice commands or touch controls~\cite{GooglePixelBuds, MobvoiEarbudsGesture}. Second, the usage scenarios for earbuds have become increasingly diverse~\cite{The5Best}. They are now commonly worn during commuting for noise cancellation, during exercise for music streaming, and in work settings for meetings or recording, with studies reporting that users in their 20s to 40s typically wear earbuds nearly every day \cite{ramage2019tinnitus} for about 50 to 60 minutes per day \cite{2023Why}. These everyday contexts provide valuable opportunities to monitor and analyze user activities in real-world environments. Third, earbuds occupy a uniquely advantageous position on the body. Situated on the head and in close proximity to facial muscles, cranial bones, the auditory canal, and major physiological structures such as the brain and large blood vessels, they can access a variety of internal signals~\cite{roddiger2022sensing}. This feature distinguishes earables from wrist- \cite{cao2021crisp} or finger-worn wearables \cite{wang2025computing} by not only measuring peripheral signals such as pulse or skin temperature but also deeper physiological cues like respiration acoustics, bone-conducted speech, heart sounds, and potentially brain-related activity. 
Fourth, their relatively stable placement on the human head helps reduce motion artifacts that are commonly encountered at the wrist, thereby improving signal quality and sensing robustness~\cite{hausamann2019ecological}. Additionally, the dual-element design (i.e., left and right earbuds) increases the information gain of sensing signals by providing multi-stream and spatially diverse inputs~\cite{cao2023heartprint,hu2024detecting}.


While commercial development has largely begun with basic earable sensing functions, such as gesture control~\cite{PixelBudsPro2, MobvoiEarbudsGesture} and heart rate monitoring~\cite{PowerbeatsPro2}, in consideration of reliability and user adoption, the academic community has been more proactive in exploring the broader sensing potential of earbuds, particularly over the past three years. We acknowledge that Röddiger \etal ~\cite{roddiger2022sensing} provided a comprehensive survey of this research field up to and including 2021, offering a taxonomy of phenomena and sensing modalities enabled by earables across various interdisciplinary communities. This survey highlighted that earables can sense a wide range of physiological signals, including motion~\cite{jiang2022earwalk, hu2024detecting, sepanosian2024boxing}, respiration~\cite{liu2024respear, hu2024breathpro}, heartbeat~\cite{butkow2023heart, cao2023heartprint}, body-conducted sounds~\cite{yang2025smarteeth,xie2022teethpass}, as well as subtler cues such as food intake~\cite{ketmalasiri2024imchew, srivastava2024bitesense}, brain activity~\cite{aziz2025unobtrusive}. They also support natural interaction through touch~\cite{zhao2024ui,yang2024maf} and voice~\cite{han2025earoe,chen2024enabling}, leveraging their proximity to the face and head. 

Building on this foundation, the pace of academic progress has accelerated dramatically since then—possibly spurred by the impact of this very survey. To our surprise, more than one hundred new papers on earable sensing have been published since 2022 in major ubiquitous computing venues\footnote{Including IMWUT, CHI, MobiCom, MobiSys, SenSys, PerCom, INFOCOM, UIST, ISWC, and associated workshops.} with an upward annual trend (as shown in Figure~\ref{fig: paperTrend}), reflecting a rapid expansion of both interest and technical capabilities within the community.
\textit{Given the scale and speed of these developments, we believe there is a clear and timely need for an updated survey to address three fundamental questions relevant to both existing researchers and newcomers to the field}: 
\begin{itemize}
    \item What kind of earable research has emerged since 2022, and how does it differ from prior work?
    \item What resources (e.g., hardware platforms and datasets) are currently available to facilitate future earable research?
    \item Is earable computing a saturated research domain, or does it remain a promising frontier? If the latter, what are the key directions for the next phase of earable research?
\end{itemize}

To answer these questions, in this survey, we conduct a comprehensive review of over one hundred recent papers on earable sensing published between 2022 and 2025. Overall, our findings highlight that, over the past three years, researchers have discovered a variety of novel sensing principles, introduced numerous new earable sensing applications, improved the performance of existing sensing tasks, and developed significant new resources to further advance research in the field. Concretely, for applications previously examined in earlier surveys, we revisit each category and analyze how recent works advance prior efforts, highlighting improvements such as enhanced accuracy, increased robustness, finer-grained contextual inference, and lightweight implementations that reduce delay and energy overhead. For emerging applications not covered before, we examine their motivations, underlying sensing principles, and the types of sensors employed, offering a structured account of how these innovations expand the functional scope of earables. In addition, we synthesize enabling resources such as hardware platforms and public datasets, and conclude with a discussion of open challenges and promising future directions for the next phase of earable research.

The rest of the paper is organized as follows: \Cref{sec: Methodology} introduces our methodology for paper retrieval and selection. Section 3 presents a taxonomy of earable sensing applications. \Cref{sec: Physiological,sec: Movement,sec: Interaction,sec: Authentication} provide an in-depth review of recent works across four major domains: physiological parameters and health (\Cref{sec: Physiological}), movement and activity (\Cref{sec: Movement}), interaction (\Cref{sec: Interaction}), and authentication and privacy (\Cref{sec: Authentication}). \Cref{sec: Resources} discusses enabling technologies and public resources such as hardware and datasets. Finally, \Cref{sec: Future} outlines future research directions, before concluding the paper in \Cref{sec:conclusion}.

\section{Methodology}
\label{sec: Methodology}

This survey aims to provide a comprehensive overview of recent research efforts on earable sensing, particularly focusing on works published between 2022 and 2025 in major ubiquitous computing venues. We adopt a structured literature review approach, combining keyword-based retrieval, manual filtering, backward citation tracing, and thematic analysis. This section outlines the steps we took in selecting and analyzing relevant works.

\subsection{Paper Retrieval}

We retrieved research articles from two major digital libraries: the ACM Digital Library (ACM-DL) and IEEE Xplore (IEEE-X), which together host a substantial portion of the literature in ubiquitous computing, human-computer interaction, and wearable systems. Our search targeted paper titles, abstracts, and metadata using an extensive list of keywords related to earable sensing. These include general terms such as \textit{earable}, \textit{hearable(s)}, \textit{ear-worn}, \textit{ear(-)mounted}, \textit{ear(-)attached}, \textit{ear(-)based}, \textit{in-ear device}, \textit{earbud(s)}, \textit{earphone(s)}, \textit{earpiece(s)}, \textit{headphone(s)}, \textit{microphone(s)}, \textit{IMU(s)}, \textit{PPG(s)}, \textit{photoplethysmography(s)}. To ensure broader coverage, we also constructed compound queries combining terms (e.g., \textit{ear-based AND authentication}, \textit{ear-based AND gesture}). 

\subsection{Selection Criteria, Filtering and Backward Chaining}

To refine the collection of papers, we applied strict inclusion criteria: 
\begin{enumerate}
    \item the paper must describe a device that is worn in, on, or around the ear; and 
    \item the sensing must take place at or near the ear, either through direct contact or close-range interaction. 
\end{enumerate}
Furthermore, we excluded works that:
\begin{enumerate}
    \item focus solely on voice and audio interfaces such as active noise cancellation, or on the technical design of audio hardware like earphones;
    \item are not peer-reviewed publications, such as patents, dissertations, or technical reports;
    \item involve larger head-worn systems, such as virtual reality headsets, which do not conform to typical earable form factors;
    \item are not written in English;
    \item do not include a functioning prototype or system evaluation.
\end{enumerate}

All papers were independently reviewed by multiple authors based on their titles, abstracts, and, when necessary, full texts to ensure consistent application of the inclusion criteria. We further conducted a backward citation search on the reference lists of the selected papers to identify additional relevant studies that were not retrieved during the initial database search. Each selected work was then annotated with key attributes, including publication year, venue, sensing modality, targeted application, sensor type, and evaluation methodology. These attributes formed the foundation for our comparative analysis, enabling us to identify emerging trends, categorize application domains, and assess both the technical depth and practical relevance of each work. Additionally, thematic coding was employed to uncover recurring design patterns and research challenges across different subfields of earable sensing. In total, we identified 111 peer-reviewed papers, which collectively form the basis for the thematic synthesis and trend analysis presented in this survey.


\section{Earable Sensing Taxonomy}
\label{sec: Taxonomy}

\begin{figure}[t]
    \includegraphics[width = 1\textwidth]{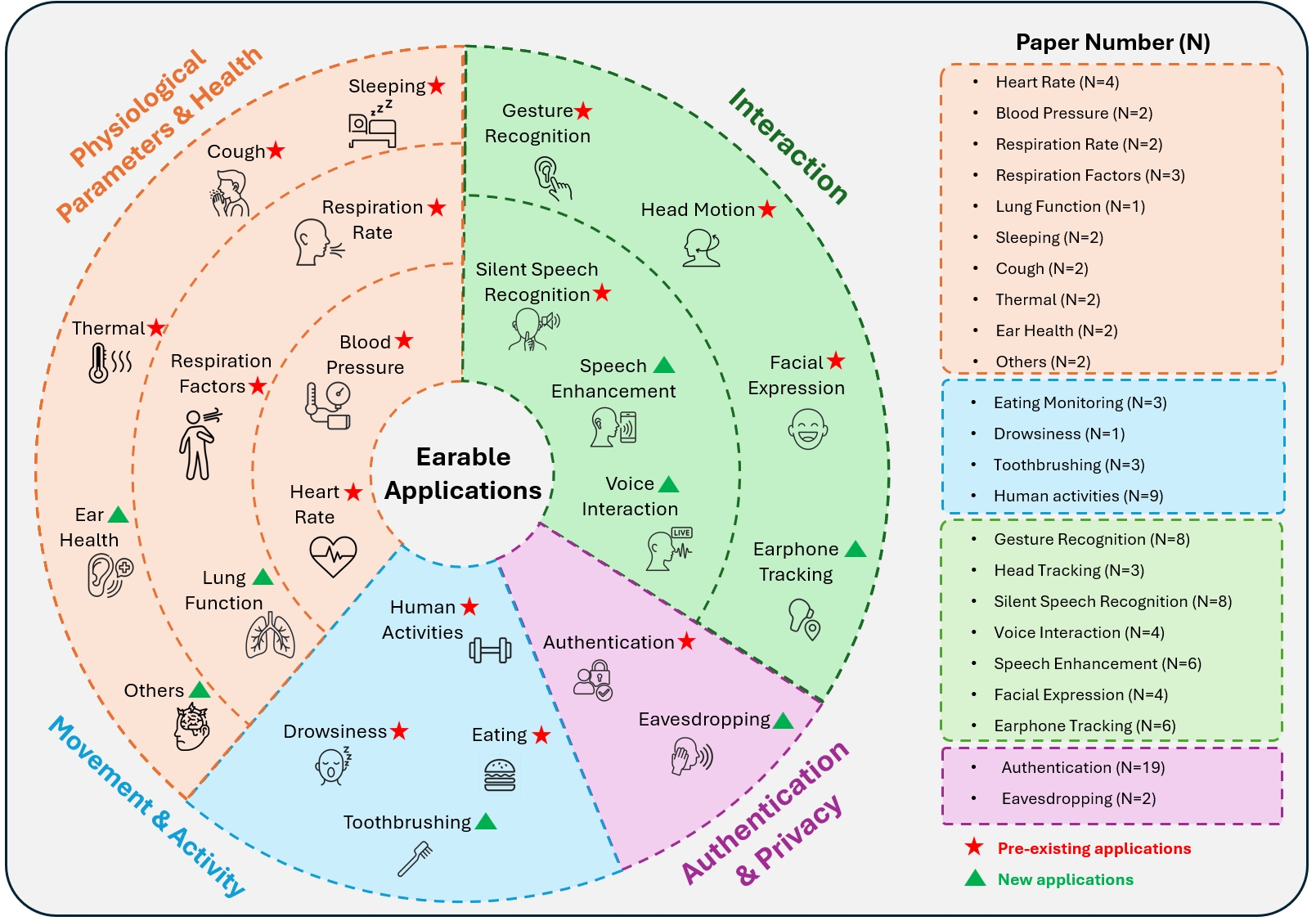}	
    \caption{Overview map of earable sensing. The map is organised by phenomena that can be captured with earables. Phenomena are clustered based on shared sensing themes or physiological systems, and grouped into four main categories: Physiological Parameters and Health (\Cref{sec: Physiological}), Movement and Activity (\Cref{sec: Movement}), Interaction (\Cref{sec: Interaction}), and Authentication and Privacy (\Cref{sec: Authentication}).}
    \label{fig: Taxonomy}
    \Description{}
\end{figure}

We follow a taxonomy structure similar to the survey \citet{roddiger2022sensing} on earable sensing and categorize applications into four primary domains: Physiological Parameters and Health, Movement and Activity, Interaction, and Authentication and Privacy, as illustrated in \Cref{fig: Taxonomy}. Each domain organizes sensing applications based on the human phenomena being monitored or enabled by earables.

To capture how the field is evolving, we further divide applications into two classes based on whether they appeared in previous earable survey~\cite{roddiger2022sensing}. The first class includes \textbf{pre-existing applications} (labeled with \textit{red pentagram} in \Cref{fig: Taxonomy}) that had been explored prior to 2022, such as gesture recognition or heart rate monitoring. For these, we analyze how recent works have addressed past limitations (such as sensitivity to noise and motion, high enrollment costs, energy inefficiency, and etc.) and identify improvements in accuracy, latency, or robustness. The second class highlights \textbf{new applications} that have emerged since the last major survey~\cite{roddiger2022sensing} (labeled with \textit{green triangles} in \Cref{fig: Taxonomy}), such as ear health detection and speech enhancement. Here, we focus on aspects such as motivation behind the application, their novel sensing principles, and unique hardware configurations.



After reviewing advancements across each domain, we conclude with a brief \textit{\textbf{Remarks}} section that distills key technical trends and deployment challenges. These remarks highlight a clear shift toward passive sensing, energy efficiency, and continual learning, along with growing emphasis on usability, cross-user generalization, and multimodal fusion. In addition to identifying these trends, we briefly highlight the remaining limitations and outline promising directions for future research. By combining technical insights with practical implications, our goal is to provide a clear and actionable overview of the evolving landscape of earable sensing.

\section{Physiological Parameters and Health}
\label{sec: Physiological}

Earable devices, ranging from everyday earphones to medical hearing aids, are emerging as compelling platforms for continuous physiological parameters and health monitoring. First, the ears are situated near rich physiological signal sources—including blood vessels, airways, and jaw muscles—enabling sensors placed in or around them to continuously capture a range of biosignals such as cardiovascular activity, respiratory patterns, and chewing or speaking behaviors with minimal attenuation. Second, current earables already incorporate microphones, IMUs, and speakers, and the same form factor can readily house additional biosensors such as PPG modules or even miniature ECG electrodes. Finally, their affordability and ubiquity make them ideal for everyday use: young adults wear earphones almost all day, from commuting to exercising, while older adults rely on hearing aids for extended periods, enabling unobtrusive, long‑term monitoring across diverse daily routines.

\begin{figure}[t]
    \includegraphics[width = 1\textwidth]{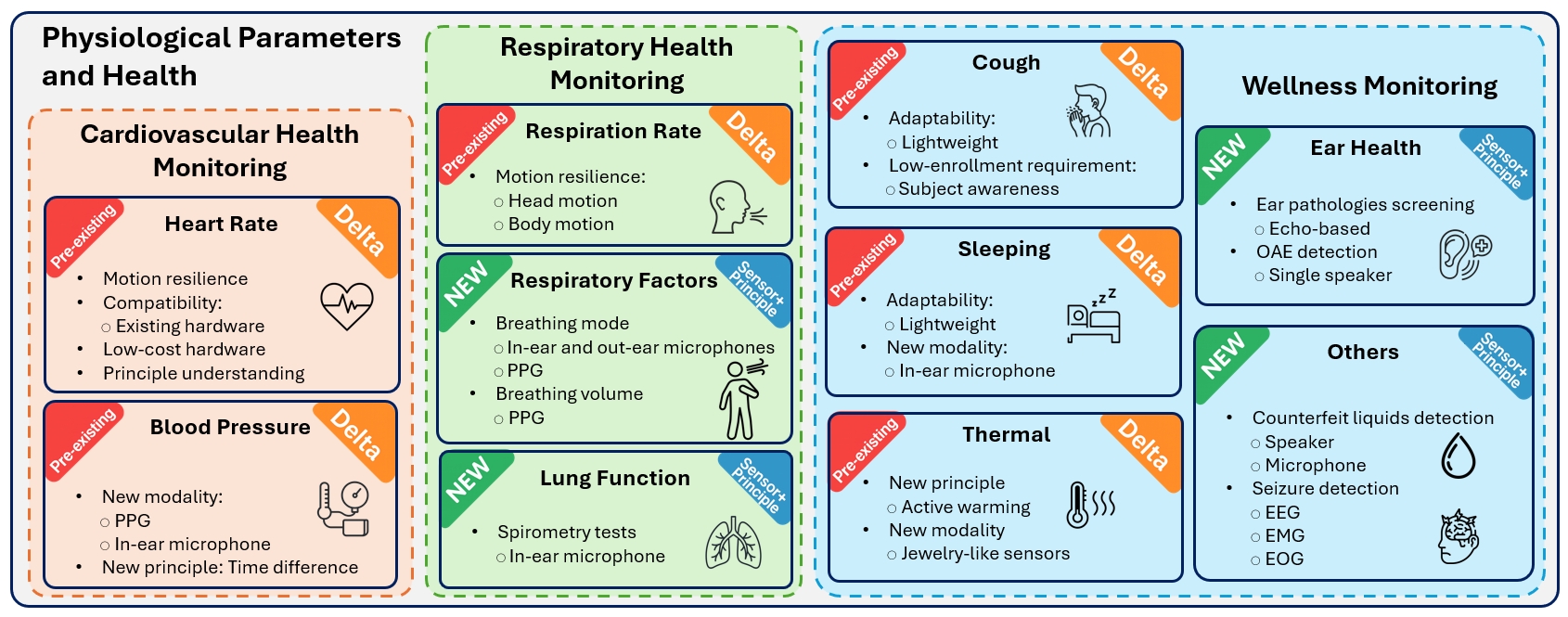}	
    \caption{Summary of recent advancements in
earable-based physiological sensing.}
    \label{fig: Physiological}
    \Description{}
\end{figure}

This section surveys recent advances in earable-based physiological sensing, focusing on three primary domains: cardiovascular health, respiratory function, and general wellness. As illustrated in Figure~\ref{fig: Physiological}, we organize applications into two main categories: (1) \textbf{Pre-existing tasks}, such as heart rate tracking, respiration rate estimation, and cough detection, which have been investigated prior to 2021; and (2) \textbf{Emerging applications}, including breathing mode recognition, ear pathology screening, and lung function analysis, which represent novel sensing opportunities uniquely enabled by the earable form factor. For each pre-existing task, we first summarize the state of the art before 2021, then discuss how recent works advance the field by addressing key limitations—such as improving motion robustness, enhancing compatibility with commercial hardware, lowering costs, or deepening physiological understanding. These incremental yet important improvements are labeled as \textbf{Delta} in the figure. For each new application, we introduce the motivation and significance of the new physiological metric, and describe the underlying sensing principles, methods, and performance. These systems often leverage innovative modalities (e.g., echo-based sensing, in-ear PPG, ear-worn EEG) or reveal new biosignal–health relationships, and are marked as \textbf{NEW} in the overview map.




\subsection{Cardiovascular Health Monitoring}
\label{sec:heart}
Continuous cardiovascular health monitoring is essential not only in clinical settings, but also for daily-life or at-home health management~\cite{sana2020wearable}. By regularly tracking indicators such as heart rate, heart rate variability, and blood pressure, both healthcare professionals and individuals can promptly detect changes in cardiac function and take timely action~\cite{mohamoud2024consumer}. Moreover, continuous cardiovascular monitoring in everyday life can enhance physical activity levels and overall cardiovascular health~\cite{petek2023consumer}.

\begin{table}[t]
\caption{Summary of earable systems for heart rate monitoring.}
\label{tab:recent_heart_rate_works}
\renewcommand{\arraystretch}{1.2}
\begin{tabular}{p{2.1cm}p{2cm}p{3.5cm}p{4.3cm}p{1.8cm}}
\hline
\textbf{System} & \textbf{Sensor} & \textbf{Application} & \textbf{Advancement} & \textbf{Performance} \\ \hline
APG \cite{fan2023apg} & Speaker\,+\,mic & HR/HRV via ultrasonic echoes & Uses only ANC hardware; motion‑resilient & 3.2 \% HR err; 2.7 \% HRV err (153 users) \\ \hline
Asclepius \cite{chen2024exploring} & Speaker (retrofit board) & PCG‑based HR/HRV & \$10 add‑on board turns wired buds into stethoscope & Cardiologist‑rated diagnostic\ PCG (30 users) \\ \hline
hEARt \cite{butkow2023heart,butkow2024evaluation} & In‑ear mic & HR under motion & Deep‑learning denoising; ANSI‑grade accuracy & MAPE < 10 \% across walk, run, rest \\ \hline
Christofferson \etal~\cite{christofferson2024production} & In‑ear mic & physiological modeling of cardiac audio & First acoustic path model for ear‑canal HR sensing & Guides sensor placement and algorithm design \\ \hline
\end{tabular}
\end{table}

\textbf{Heart Rate.} Over the past few years, hearable‑based cardiovascular research has focused chiefly on heart rate metrics, heart rate (HR) and heart rate variability (HRV), while increasingly emphasizing three goals: (i) compatibility with ubiquitous, low‑cost wired earphones that contain only speakers~\cite{chen2024exploring}, (ii) robust performance in challenging real‑world conditions~\cite{fan2023apg,butkow2023heart,butkow2024evaluation}, and (iii) a deeper understanding of the underlying physiology~\cite{christofferson2024production}, as summarized in \cref{tab:recent_heart_rate_works}. Fan \etal~\cite{fan2023apg} introduced APG, which injects low‑intensity ultrasonic probes through off‑the‑shelf ANC headphones and measures echo modulations with the built‑in feedback microphones, achieving median errors of 3.2\% for HR and 2.7\% for HRV across 153 participants who wore the devices while sitting, walking, and running, even with music playing in the background. Building on the idea of using only the speaker hardware, Chen \etal~\cite{chen2024exploring}’s Asclepius system retrofits a < US \$10 plug‑in board that turns commodity wired earphones into an electronic stethoscope; signal‑processing and learning pipelines recover high‑fidelity phonocardiograms (PCG) from the ear canal, and blind evaluations with five cardiologists on data from 30 volunteers confirmed diagnostic quality heart sounds and accurate HR/HRV extraction. To improve robustness in motion‑rich environments, Butkow \etal~\cite{butkow2023heart} proposed hEARt, a deep‑learning framework that denoises in‑ear microphone signals and estimates HR. It achieved mean absolute errors (MAE) of 3.0 ± 3.0 BPM, 8.1 ± 6.7 BPM, 11.2 ± 9.2 BPM, and 9.4 ± 7.0 BPM during stationary, walking, running, and speaking conditions, respectively. The extension of this work further~\cite{butkow2024evaluation} offers a broader evaluation and shows that the method outperforms in‑ear PPG under identical motion conditions, achieving MAEs of 1.9 BPM when sedentary, 6.8 BPM while walking, and 13.2 BPM while running across 15 subjects and nine activities. Finally, Christofferson \etal~\cite{christofferson2024production} conducted the first systematic study of how cardiac vibrations propagate to the ear canal, i.e., how in-ear cardiac audio is produced, combining acoustic modeling with in‑situ measurements to quantify frequency content, attenuation paths, and inter‑subject variability; their findings supply the physiological grounding that guides sensor placement and algorithm design for future in-ear acoustic based cardiovascular systems. Together, these works illustrate the field’s shift from proof‑of‑concept HR sensing toward low‑cost hardware reuse, motion‑resilient algorithms, and foundational physiological insight—key steps toward ubiquitous, hearable‑based cardiovascular monitoring.

\textbf{Blood Pressure.} Parallel research streams have extended hearable sensing beyond heart rate metrics by investigating blood pressure (BP) estimation~\cite{balaji2023stereo,zhao2024hearbp}. Balaji \etal~\cite{balaji2023stereo} proposed Stereo-BP, an earable system for non-invasive BP monitoring that relies on photoplethysmography (PPG) signals collected from both ears simultaneously. By measuring pulse arrival time differences between the left and right ear PPG waveforms, the system demonstrated feasibility for estimating systolic and diastolic BP, achieving MAEs of 3.97 mmHg and 3.83 mmHg, respectively, in a preliminary study with 20 participants. Meanwhile, Zhao \etal~\cite{zhao2024hearbp} introduced HearBP, an in-ear BP monitoring system based on bone-conducted heart sounds. By recording heart sounds via microphones in both ear canals, the system targets the time interval between the first and second heart sounds as a primary indicator for BP estimation. A U-net autoencoder-decoder removes ambient noise to isolate clean heart sounds, while feature extraction leverages Shannon energy and energy-entropy ratio, followed by principal component analysis for dimension reduction. The final BP prediction uses a dendritic neural regression model, achieving an average error of 0.97 mmHg for diastolic and 1.61 mmHg for systolic BP, with standard deviations of 3.13 and 3.56 mmHg, respectively, meeting FDA’s AAMI protocol in tests with 41 participants. \textit{\textbf{Advancement over Prior Work:} These recent efforts mark a significant step toward expanding the clinical utility of hearables by demonstrating the feasibility of blood pressure estimation using only non-intrusive, in-ear sensors such as PPG or microphones. By eliminating the need for traditional cuff-based or multi-site measurements, these systems highlight the field’s growing emphasis on comfort, wearability, and continuous monitoring in daily life—key enablers for broader, real-world cardiovascular assessment.}

\textbf{Remarks.}  Recent studies show that commodity earphones can (i) deliver sub‑3 \% HR/HRV error and single‑digit ‑mmHg BP error without cuffs, (ii) retain accuracy while users walk, run, or listen to music, and (iii) ground sensor design in new models of how cardiac vibrations reach the ear canal.  To move from promising prototypes to everyday devices, several challenges remain:
\begin{itemize}
    \item \textbf{Motion Robustness.} First, reliably handling motion artifacts across diverse activities remains a major challenge, with performance under motion still having room to improve, and requiring smoother transitions between everyday conditions.
    \item \textbf{Hardware Compatibility.} Second, although most prototypes employ sensors similar to those found in commercial earphones, the hardware gap between these prototypes and real‑world devices has not been fully investigated.
    \item \textbf{Signal Interference \& Scope Expansion.} Third, the primary functions of earphones—music playback and speech—can interfere with physiological signal acquisition, underscoring the need for robust interference mitigation. Beyond HR, HRV, and BP, expanding monitoring to include additional cardiovascular factors (e.g., cardiac timing intervals, arrhythmia detection, or vascular stiffness) may also be a promising direction.
    \item \textbf{Clinical Validation.} Finally, more extensive clinical evaluations are necessary—especially those involving data collection from patients in longitudinal, in-the-wild studies—to validate system robustness, reliability, and clinical relevance in real-world medical and lifestyle contexts.
\end{itemize}

\subsection{Respiratory Health Monitoring}
\label{sec:breath}
Continuous respiratory health monitoring is essential not only in clinical settings, but also for everyday scenarios like sports training, sleep tracking, and overall wellness~\cite{vitazkova2024advances,massaroni2019contact}. By closely observing indicators such as breathing rate, airflow patterns, and other aspects of lung function, both healthcare professionals and individuals can detect subtle changes early and respond promptly~\cite{vitazkova2024advances}. Moreover, it can provide insights into exercise efficiency, sleep quality, and promote overall wellbeing~\cite{massaroni2019contact}. Building on the unique advantages of hearable sensing, 
recent research in hearable-based respiratory health monitoring has primarily focused on several key physiological metrics, including respiration rate (RR)~\cite{liu2024respear,ahmed2023remote,romero2024optibreathe}, breathing modes~\cite{romero2024optibreathe,hu2024breathpro}, tidal volume~\cite{hu2024breathpro}, and lung function assessment~\cite{xie2023earspiro}. A summary of these recent advancements is provided in \cref{tab:earable_respiratory_advances}.


\begin{table}[t]
\caption{Summary of earable systems for respiratory monitoring.}
\label{tab:earable_respiratory_advances}
\renewcommand{\arraystretch}{1.2}
\begin{tabular}{p{2.4cm}p{1.5cm}p{3.2cm}p{4.3cm}p{2.2cm}}
\hline
\textbf{System} & \textbf{Sensor} & \textbf{Application} & \textbf{Advancement} & \textbf{Performance} \\ \hline
Ahmed \etal~\cite{ahmed2023remote} & IMU, Mic & Multimodal RR detection (head‑motion robust) & Dynamic modality switching; balances accuracy and data retention & MAE $<$ 2 BPM, 75.1 \% session retention \\ \hline
RespEar~\cite{liu2024respear} & Mic & RR tracking across rest \& activity (RSA/LRC coupling) & First to generalise to sedentary + active states with in-ear mic & MAE 1.48 BPM (rest), 2.28 BPM (active) \\ \hline
BreathPro~\cite{hu2024breathpro} & Mic (in + out) & Breathing‑mode detection during running & First nasal/oral phase recognition in motion & 98.5 \% Acc (lab), 89.9 \% Acc (wild) \\ \hline
OptiBreathe~\cite{romero2024optibreathe} & PPG & RR, phase, tidal‑volume estimation & First multi‑parameter respiratory sensing via in‑ear PPG & MAE 1.96 BPM (RR); 17 \% MAPE (tidal volume) \\ \hline
EarSpiro~\cite{xie2023earspiro} & Mic & Full flow–volume curves (lung function) & First earable spirometry covering insp + exp phases & MAE 0.20 L/s (exp), 0.42 L/s (insp); 0.94 correlation \\ \hline
\end{tabular}
\end{table}

\textbf{Respiration Rate.} Recent efforts in RR monitoring have increasingly focused on more challenging scenarios, particularly under motion-rich conditions. These studies aim to ensure robust and accurate RR estimation during everyday activities such as stationary conditions (with head movements), walking, running, or exercising, where motion artifacts can significantly degrade signal quality. Ahmed \etal ~\cite{ahmed2023remote} proposed a multimodal RR monitoring system using motion and acoustic sensors in commercial earables (Samsung Galaxy Buds Pro) to track breathing in stationary conditions (with head movements). The system combines a motion-based signal processing pipeline with an audio-based machine learning model, switching between them as needed to ensure reliability, i.e., balancing the accuracy and data retention. In a week-long study with 30 participants and over 2500 sessions, the system achieved a MAE below 2 breaths per minute (BPM) and a 75.1\% session retention rate—outperforming prior unimodal approaches while remaining lightweight enough for on-device use. Liu \etal ~\cite{liu2024respear} introduced RespEar, an earable system that leverages in-ear microphones to monitor RR under both sedentary and active conditions by leveraging physiological couplings—Respiratory Sinus Arrhythmia during rest and Locomotor Respiratory Coupling during movement. Evaluated on 18 subjects across eight activities, RespEar achieved MAE of 1.48 BPM in sedentary states and 2.28 BPM during active scenarios, making it the first earable system to demonstrate robust performance across a wide range of daily activities, even under motion-rich conditions.

\textbf{New Respiratory Factors.} Beyond respiratory rate estimation, recent work has also begun exploring richer aspects of respiratory health that had not been previously investigated. Hu \etal ~\cite{hu2024breathpro} proposed BreathPro, the first earable system for recognizing breathing modes (nasal/oral inhalation and exhalation) during running. Using in-ear and out-ear microphones, the system captures respiratory sounds and applies personalized noise reduction, feature extraction, and KNN-based classification. Evaluated on 25 participants, BreathPro achieved 98.52\% phase-level accuracy in lab conditions and 89.93\% accuracy during in-the-wild running, demonstrating robust real-time breathing mode detection under varying ambient noise. Romero \etal ~\cite{romero2024optibreathe} proposed OptiBreathe, an earable-based system that uses in-ear PPG signals to continuously monitor three key respiratory metrics: RR, breathing phases (inhalation and exhalation), and tidal volume under stationary. The system applies lightweight signal processing techniques to extract respiratory-induced variations in the PPG signal, including baseline wander, amplitude changes, and frequency shifts. Evaluated on 11 participants with medical-grade spirometry as ground truth, OptiBreathe achieved a MAE of 1.96 BPM for RR, 0.48 seconds and 0.14 for inspiratory time and inhalation-to-exhalation ratio respectively, and a mean absolute percentage error (MAPE) of 17\% for tidal volume. These results demonstrate the potential of in-ear PPG for comprehensive respiratory monitoring.

\textbf{Lung Function.} In addition to tracking respiratory health metrics, recent studies have started to investigate the novel potential of earables in lung function assessment. Xie \etal ~\cite{xie2023earspiro} proposed EarSpiro, the first earphone-based system capable of reconstructing complete flow–volume (F-V) curves—including both expiratory and inspiratory phases—for lung function assessment. Unlike prior mobile solutions that only estimate discrete lung indices, EarSpiro captures airflow sounds during spirometry tests using in-ear microphones and translates them into detailed F-V curves. The system leverages a combination of convolutional and recurrent neural networks to model the complex relationship between sound and airflow speed, and employs a clustering-based segmentation algorithm to isolate weak inspiratory signals. It also supports the use of household funnel-like objects as improvised mouthpieces via transfer learning, using a few personalized lung indices to fine-tune predictions. Evaluated on 60 subjects, EarSpiro achieved mean errors of 0.20 L/s and 0.42 L/s for expiratory and inspiratory flow rate estimation, a correlation coefficient of 0.94 for F-V curves, and an average error of 7.3\% for four key lung function indices—demonstrating its promise as a low-cost, accessible tool for at-home lung function monitoring.

\textbf{Remarks.}  Collectively, the latest systems show that (i) commodity in‑ear sensors can now track RR with $\le\!2$ BPM error even during motion, (ii) richer respiratory factors—breathing modes, tidal volume, full flow–volume curves—are feasible with microphones or PPG alone, and (iii) learning‑based models grounded in respiratory physiology can translate these signals into clinically meaningful metrics.  To push hearable respiratory monitoring from feasibility studies to everyday utility, several complementary directions remain:
\begin{itemize}
    \item \textbf{User‑Tailored Metrics.}  Future systems should expose \emph{application‑specific outputs}—for example, breathing efficiency or ventilation patterns for endurance athletes—so that the same core sensing pipeline can deliver the most relevant information to different user groups.
    
    \item \textbf{Broad Generalizability.}  To ensure that a single model serves everyone fairly, algorithms need to be trained and validated on more heterogeneous cohorts, including healthy individuals, patients with respiratory disorders, and users spanning a wide range of ages and fitness levels.
    
    \item \textbf{Lightweight Personalization.}  Finally, even with a broadly trained model, small per‑user differences in ear anatomy, breathing style, and ear‑tip fit can degrade accuracy.  Techniques such as domain adaptation or a brief, on‑device calibration step can provide \emph{personalized refinements} without imposing a heavy data‑collection burden.
\end{itemize}

\subsection{Wellness Monitoring}
\label{sec:wellness}
Beyond cardiovascular and respiratory monitoring, recent research has leveraged earables to track a wider range of wellness‑related factors—including coughing~\cite{wang2022hearcough,zhang2023earcough}, sleep quality~\cite{christofferson2022sleep,han2024earsleep}, and thermal regulation~\cite{xue2024design,knierim2024warmth}—as well as newer applications such as ear‑health assessment~\cite{jin2022earhealth,chan2023wireless} and other emerging use cases~\cite{li2024asliquid,aziz2025unobtrusive}.

\textbf{Cough.} Recent work has leveraged earables for continuous cough monitoring, aiming to support airway-related health assessment in everyday settings. \textit{\textbf{Advancement over Prior Work:} The latest systems push two key frontiers: (i) running lightweight, energy‑efficient networks entirely on earbud‑class microcontrollers to preserve privacy and battery life, and (ii) adding ``subject awareness'' so the device can tell the wearer’s coughs from those of nearby people, boosting clinical relevance.} Wang \etal ~\cite{wang2022hearcough} proposed HearCough, a system that enables cough event detection using in-ear microphones embedded in commercial active noise cancellation (ANC) earphones. They designed a lightweight end-to-end neural network, Tiny-COUNET, which runs entirely on-device without requiring cloud offloading. Deployed on a microcontroller used in commodity earbuds, HearCough achieved 90.0\% accuracy and 89.5\% F1-score in a field study with 8 patients, consuming only 5.2 mW of additional power—demonstrating feasibility for low-cost, privacy-preserving, and continuous cough detection. Expanding on this, Zhang \etal ~\cite{zhang2023earcough} introduced EarCough, which focuses specifically on subject-aware cough detection—distinguishing the user’s own coughs from those in the environment. The proposed model, EarCoughNet, uses dual-channel input from hybrid ANC microphones (feedforward and feedback) to enhance subject awareness. Trained and evaluated on a newly collected multimodal dataset (audio + motion), EarCough achieved 95.4\% accuracy and 92.9\% F1-score, all with a compact 385 kB model suitable for deployment on modern earbud microcontrollers. This work highlights the importance of personalized cough detection and the potential of dual-microphone signal differentiation for real-world robustness.

\textbf{Sleeping.} Moreover, recent research has demonstrated the feasibility of using in-ear acoustic sensing for sleep quality assessment. \textit{\textbf{Advancement over Prior Work:} The newest systems show two clear advances: (i) running compact neural models entirely on ear‑level or phone hardware to deliver truly unobtrusive, at‑home sleep analytics, and (ii) moving beyond snore detection to fuse body‑conducted and ambient sounds for fine‑grained, multi‑stage sleep characterization.} Christofferson \etal ~\cite{christofferson2022sleep} proposed an approach using ANC earphones to classify sounds associated with disordered sleep, such as snoring, teeth grinding, and movement. Their system uses both internal and external microphones to capture dual-channel audio and employs a lightweight deep learning model based on a modified temporal shift module for low-power inference. Tested on 20 participants with 8 sleep-relevant sound categories, the system achieved 91.0\% accuracy and an F1-score of 0.845, all while being compact enough to run on smartphones or earbuds themselves. Han \etal ~\cite{han2024earsleep} made a step further to introduce EarSleep, a system that leverages in-ear microphones to capture body-conducted sounds generated by various sleep-related activities—such as body movements, snoring, coughing, heartbeat, and breathing. These signals are processed through a dual-stream deep learning framework to extract both physical and physiological features for fine-grained sleep stage detection. In a three-month study with 18 participants, EarSleep outperformed prior wearable systems by 7.12\% in average precision and 9.32\% in average recall, demonstrating robust, interpretable, and non-invasive sleep monitoring using a single sensing modality.

\textbf{Thermal.} In addition to sensing physiological signals, recent studies have also explored how earables can support thermal comfort and longitudinal body temperature monitoring. \textit{\textbf{Advancement over Prior Work:} The newest prototypes push two complementary frontiers: (i) thermo‑active earwear that can actively warm or cool the wearer on demand, and (ii) jewelry‑like sensors that unobtrusively log skin temperature for weeks on a single charge—together pointing toward seamless, day‑long thermal well‑being.} Knierim \etal ~\cite{knierim2024warmth} introduced Warmth on Demand, a thermoactive headphone prototype designed to deliver localized heating around the ears using Peltier elements embedded in over-ear headphones. In a controlled study, participants reported significantly improved thermal sensation and comfort in both the ear region and whole body, particularly in cold office environments. Although thermal effects were mostly local, findings suggest that wearable thermal systems could support user comfort and even productivity in workplace settings. On the other hand, Xue \etal ~\cite{xue2024design} developed Thermal Earring, a low-power wireless earring for continuous earlobe temperature monitoring. Designed to resemble traditional jewelry, the smart earring includes dual temperature sensors for ambient and skin measurements, runs on 14.4uW, and lasts up to 28 days on a single charge. Its unobtrusive form factor enables applications such as fever detection, stress monitoring, and activity tracking, offering a fashionable and practical platform for long-term wellness sensing.

\begin{table}[t]
\caption{Summary of earable systems for new applications in wellness monitoring.}
\label{tab:earable_wellness_advances}
\renewcommand{\arraystretch}{1.2}
\begin{tabular}{p{2.1cm}p{2cm}p{3.5cm}p{4.3cm}p{1.8cm}}
\hline
\textbf{System} & \textbf{Sensor} & \textbf{Application} & \textbf{Advancement} & \textbf{Performance} \\ \hline
\multicolumn{5}{c}{\textbf{Ear‑health monitoring}} \\ \hline
EarHealth \cite{jin2022earhealth} & Speaker\,+\,mic & Screen ruptured eardrum, otitis media, ear‑wax blockage & First earphone system to detect multiple ear pathologies outside clinics & 82.6 \% Acc \\ \hline
OAEbuds \cite{chan2023wireless} & Single speaker\,+\,mic & Otoacoustic‑emission hearing screening & FDA‑comparable accuracy at low cost, fully wireless & 100 \% Sensitivity \\  & & & & 89.7 \% Specificity \\ \hline
\multicolumn{5}{c}{\textbf{Other emerging applications}} \\ \hline
ASLiquid \cite{li2024asliquid} & Speaker\,+\,mic & Counterfeit‑liquid detection via acoustic resonance & First liquid‑authenticity sensing with commodity earphones & 95–99.3 \% F$_1$ \\ \hline
EarSD \cite{aziz2025unobtrusive} & EEG/EMG/EOG electrodes & Continuous seizure detection & First ear‑mounted platform offering near‑clinical seizure monitoring & 97.9 \% Acc \\ \hline
\end{tabular}
\end{table}

\textbf{New Applications—Ear Health.} As summarized in \cref{tab:earable_wellness_advances}, emerging earable research now targets specialized audiological tasks. Recent work is pushing earables beyond general wellness sensing toward clinical‑grade ear‑health assessment, enabling low‑cost, in‑situ screening of common ear pathologies and cochlear function. Jin \etal ~\cite{jin2022earhealth} proposed EarHealth, an earphone-based acoustic sensing system that identifies three prevalent ear conditions—ruptured eardrum, otitis media, and earwax blockage—by analyzing the recorded echoes evoked by a chirp sound stimulus in the ear canal. The system extracts features related to eardrum mobility and ear canal geometry, and applies a multi-view deep learning model for classification. Evaluated on 92 participants across four conditions, EarHealth achieved 82.6\% overall accuracy, demonstrating the feasibility of using commercial earphones for convenient, continuous ear health screening in non-clinical settings. Complementing this, Chan \etal ~\cite{chan2023wireless} introduced OAEbuds, the first low-cost wireless earbud system for hearing screening via detection of otoacoustic emissions (OAEs)—faint sounds generated by a healthy cochlea. Their design combines frequency-modulated chirps and wideband pulses to separate OAEs from in-ear echoes using only a single low-cost speaker. In clinical tests involving 50 ears across two healthcare sites, OAEbuds achieved 100\% sensitivity and 89.7\% specificity, comparable to an \$8000 FDA-cleared device, while remaining fully open-source and affordable for widespread use.

\textbf{New Applications—Others.} Moving beyond conventional vital‑sign monitoring, a new wave of studies is repurposing earables for emerging use cases that span consumer wellness, product safety, and even novel clinical scenarios—ranging from counterfeit‑liquid detection to unobtrusive seizure alerts—thereby greatly broadening their role in everyday life, as summarized in \cref{tab:earable_wellness_advances}. Li \etal ~\cite{li2024asliquid} introduced ASLiquid, the first system that uses speakers and microphones on commercial off-the-shelf earphones to detect counterfeit liquids. By leveraging acoustic resonance properties within a liquid-container system, ASLiquid identifies subtle differences in liquid density and solute composition. Through advanced signal processing and a variational autoencoder-based anomaly detector, the system achieved 95–99.25\% F1 scores across seven real-world liquid fraud scenarios—including alcohol, milk, and perfume—demonstrating that earphones can serve as ubiquitous, low-cost sensors for product authenticity verification. In another novel direction, Aziz \etal ~\cite{aziz2025unobtrusive} developed EarSD, a lightweight, ear-worn system for continuous epileptic seizure detection. EarSD captures electroencephalography (EEG), electromyography (EMG), and electrooculography (EOG) signals from electrodes placed behind the ears and uses advanced decomposition and machine learning techniques to detect seizure events in real-time. In both lab and hospital trials involving 32 patients, the system achieved up to 97.9\% accuracy, offering a socially acceptable and unobtrusive alternative to conventional video-EEG setups. This work highlights the potential for integrating earables into neurological disorder management beyond traditional hospital settings.

\textbf{Remarks.}  Collectively, the latest studies show that commodity earables can now (i) run on‑device, privacy‑preserving models for cough and sleep analysis, (ii) deliver both health and comfort functions—such as continuous temperature logging or on‑demand ear warming—and (iii) extend into entirely new domains, from clinical ear‑pathology screening to counterfeit‑liquid detection and seizure alerts.  These advances move earables from single‑purpose sensors to versatile wellness platforms, yet several challenges remain:
\begin{itemize}
    \item \textbf{Holistic Multi‑Symptom Sensing.}  First, many existing solutions focus on single‑symptom or single‑event detection (e.g., coughs, snoring); future work could explore multi‑symptom integration to provide a more holistic understanding of user health and behavioral context.
    
    \item \textbf{Longitudinal Validation.}  Second, longitudinal studies with diverse user populations—including different age groups, occupations, and health conditions—are needed to validate system robustness, usability, and long‑term value in everyday life.
    
    \item \textbf{Multimodal Fusion \& Edge Efficiency.}  Third, the fusion of multimodal sensing, such as combining audio, motion, and temperature signals, offers promising opportunities for richer inference but also raises new challenges in synchronization, privacy, and on‑device processing efficiency.

    \item \textbf{Personalized Adaptation.}  Finally, personalization and adaptability remain largely underexplored: systems that adapt to an individual's baseline behavior and environmental context could significantly improve accuracy and user trust.
\end{itemize}
\section{Movement and Activity}
\label{sec: Movement}

\begin{figure}[t]
    \includegraphics[width = 1\textwidth]{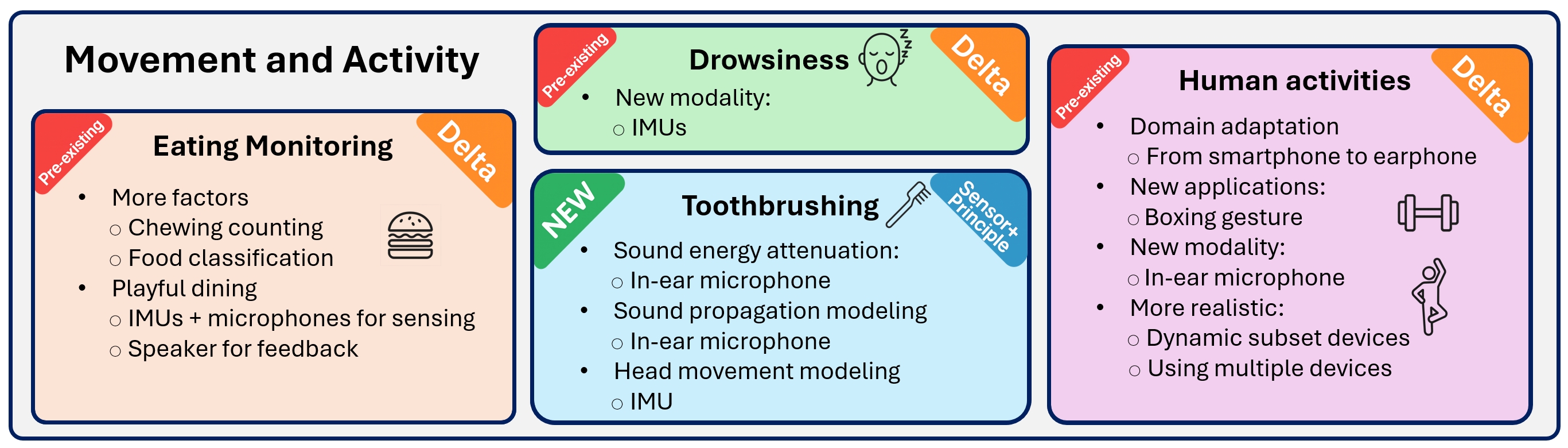}	
    \caption{Summary of recent advancements in
earable-based movement and activity sensing.}
    \label{fig: Movement}
    \Description{}
\end{figure}

Earable devices have shown growing potential in capturing diverse aspects of human movement and activity. As shown in Figure~\ref{fig: Movement}, this section reviews recent progress in earable applications for movement and activity sensing.  Prior work has explored areas such as eating monitoring, drowsiness detection, and general human activity recognition. In eating monitoring, recent studies go beyond simple food intake detection to include chewing counting, food classification, and even interactive dining experiences, enabled by integrating IMUs and microphones for sensing and using speakers for real-time feedback. Drowsiness detection has now begun to leverage IMUs embedded in earables as a new modality. For general human activity recognition, recent efforts focus on domain adaptation and expanding the range of detectable activities, such as boxing gestures. Beyond these extended directions, toothbrushing monitoring emerges as a new and unique application for earables. It introduces novel sensing principles such as in-ear acoustic attenuation and propagation modeling, coupled with IMU-based head motion tracking, opening up possibilities for personalized, at-home oral health monitoring. These advances reflect a broader shift toward fine-grained behavior tracking in daily life, with improved sensing realism and system adaptability across diverse user actions. 


\subsection{Eating Monitoring}
\label{sec:eating}
Research has explored how earables can unobtrusively monitor eating behaviors, with the dual goals of promoting healthy eating habits and enhancing dining experiences. These systems leverage ear-worn sensors to capture head and jaw movements associated with chewing and drinking, offering a discreet and socially acceptable alternative to traditional methods as reviewed in~\cite{roddiger2022sensing}.
Recent work has quickly progressed from basic chew detection to richer, user‑centred capabilities: today’s systems (1) turn raw jaw or head motions into quantitative meal metrics such as chew count and pace, (2) combine motion and audio cues to recognise food type, texture, and nutritional characteristics, and (3) even generate real‑time, playful audio feedback that heightens dining enjoyment and mindfulness.

\begin{table}[t]
\caption{Summary of earable systems for eating monitoring.}
\label{tab:earable_eating_advances}
\renewcommand{\arraystretch}{1.2}
\begin{tabular}{p{2.1cm}p{2cm}p{3.5cm}p{4.3cm}p{1.8cm}}
\hline
\textbf{System} & \textbf{Sensor} & \textbf{Application} & \textbf{Advancement} & \textbf{Performance} \\ \hline
IMChew \cite{ketmalasiri2024imchew} & IMU & Chew count estimation & First to convert binary chew detection into quantitative chew frequency & 91 \% Acc, 9.5 \% MAPE \\ \hline
BiteSense \cite{srivastava2024bitesense} & IMU & Food type recognition + bite metrics & Fine‑grained dietary profiling across diverse ages and contexts & F$_1$ = 0.86; 8–12 \%↑ chew‑episode detection \\ \hline
GustosonicSense \cite{wang2024gustosonicsense} & Mic + IMU & Real‑time audio feedback for playful dining & Shifts from health monitoring to multisensory experiential augmentation & In‑the‑wild study: ↑ engagement, enjoyment, mindfulness \\ \hline
\end{tabular}
\end{table}


Ketmalasiri \etal ~\cite{ketmalasiri2024imchew} proposed IMChew, an earphone-based system that uses inertial measurement units (IMUs) to not only detect but also count chewing activity. IMChew consists of two key components: a chewing detector and a chewing counter. The detector uses a combination of time- and frequency-domain features with machine learning classifiers to distinguish chewing from non-chewing activity, while the counter estimates chewing frequency within detected chewing episodes. Evaluated using leave-one-subject-out cross-validation on data from eight participants, IMChew achieved 91\% accuracy and F1-score for chewing detection and a MAPE of 9.5\% for chewing count estimation—highlighting its feasibility for personalized, unobtrusive eating monitoring in daily life. Then, Srivastava \etal ~\cite{srivastava2024bitesense} proposed BiteSense, an earable-based system that leverages IMUs in commercial earphones to perform comprehensive eating behavior assessment. Beyond detecting chewing activity, BiteSense enables food classification based on properties such as state (solid/liquid/semi-liquid), texture (crunchy/soft), cooking method, and nutritional value. It also extracts behavioral metrics like bite count, chewing force, chewing duration, and eating pace, allowing inference of high-level patterns such as meal type and potential over- or undereating. The system uses a transformer-based temporal model for robust performance across diverse users and contexts. Evaluated on 38 participants aged 6–80 in both semi-controlled and real-world scenarios, BiteSense achieved an F1-score of 0.86 for food classification and outperformed prior work like EarBit and IMChew in chewing episode detection by 8–12\%. This work highlights the potential of earables for unobtrusive, fine-grained, and long-term dietary monitoring. In a complementary direction, Wang \etal ~\cite{wang2024gustosonicsense} introduced GustosonicSense, a system that uses earbuds with built-in microphones and IMUs to sense eating and drinking actions and trigger playful, real-time audio feedback. Unlike systems focused solely on health monitoring, GustosonicSense explores the experiential and playful aspects of eating by mapping different food textures to custom soundscapes (e.g., crunchy sounds, classical music). In an in-the-wild study, users reported enhanced engagement, enjoyment, and mindfulness during meals, with the system supporting stimulation, hedonism, and reflexivity in dining. This work illustrates a novel use of earables in augmenting eating as a multisensory, playful experience.

\textbf{Remarks.} Recent studies have advanced earable eating monitoring from simple chew detection to: (i) quantitative meal metrics such as chew count and pace, (ii) fine‑grained dietary profiling that recognises food type, and (iii) experiential augmentation that enriches dining through real‑time audio feedback. To move from feasibility to everyday utility, several challenges remain:
\begin{itemize}
    \item \textbf{Nutrition‑Centric Analytics.} First, future work could explore richer eating‑behavior analysis—such as nutrition analysis and management—to provide more detailed dietary monitoring and feedback.
    
    \item \textbf{Contextual Robustness.} Second, improving robustness across diverse eating contexts—including different postures, environments, and habits—will be key to ensuring reliable use in naturalistic settings.
    
    \item \textbf{Health–Experience Integration.} Third, integrating health‑focused and experiential goals presents an exciting direction: systems could simultaneously encourage healthy eating behaviors while enhancing user enjoyment and mindfulness.
\end{itemize}

\subsection{Drowsiness}
\label{sec:drowsiness}
Fatigue and drowsiness detection are critical for safety and productivity in contexts such as long-distance driving, night shifts, or high-attention tasks. \textit{\textbf{Advancement over Prior Work:} Recent work takes a step toward leveraging ubiquitous sensing modalities in earables for this task.} Brown \etal ~\cite{brown2023yawning} proposed a solution for yawning detection using IMUs in earphones, offering a privacy-preserving, low-cost, and portable alternative to traditional vision- or vehicle-based systems. The system leverages head and jaw motion captured by a 6-axis IMU in standard earphones to distinguish yawns from other daily activities such as walking, talking, or eating. They designed five deep learning models and three traditional machine learning classifiers based on time- and frequency-domain features. Evaluated on a dataset collected from 10 participants as well as a public benchmark, the system achieved F1 scores of up to 0.90 on the collected dataset and 0.71 on the public dataset, validating its feasibility for robust drowsiness monitoring.

\textbf{Remarks.} Recent earable drowsiness studies have proven IMU‑based yawning detection feasible. Future work on earable-based drowsiness detection could focus on several key areas:
\begin{itemize}
    \item \textbf{Multi‑Signal Fusion.} First, combining yawning detection with additional physiological or behavioral cues—such as head nodding, microsleeps, or breathing changes—could improve accuracy and robustness.
    
    \item \textbf{Personalized Modeling.} Second, developing models that adapt to each user’s motion and physiology may enhance real‑world performance.
    
    \item \textbf{Longitudinal Field Evaluation.} Third, expanding studies to longitudinal, in‑the‑wild scenarios—night‑time driving, shift work, etc.—is essential for assessing practical usability and reliability.
    
    \item \textbf{Real‑Time Intervention.} Lastly, integrating timely feedback or alert mechanisms would make such systems more actionable in safety‑critical environments.
\end{itemize}

\subsection{Toothbrushing}

Maintaining proper oral hygiene is essential for overall health, with toothbrushing being the most common and effective method of plaque removal. 
Many prior studies have explored using wrist-worn IMU sensors, such as smartwatches, to track brushing activity. However, these approaches are susceptible to noise from general hand movements and non-brushing activities, limiting their accuracy in practical scenarios \cite{huang_toothbrushing_2016,luo_brush_2019,akther_mteeth_2021}. Recently, earable sensing has been leveraged to monitor daily toothbrushing activity in real-world scenarios due to the ubiquitous presence of sensors and the location close to the mouth.

BrushBuds \cite{yang2024brushbuds} addressed the limitations of wrist-worn IMUs by directly using the motion sensors in earphones. By utilizing accelerometers and gyroscopes, the system captured subtle head and jaw movements during brushing, providing a more stable sensing method. The result shows BrushBuds can achieve a detection accuracy of 84.3\% for six different mouth regions. While BrushBuds successfully distinguished different brushing regions, it struggled to differentiate between surfaces of the same tooth due to the similar head movements, limiting its precision. To overcome the limitations of IMU-based tracking, ToothFairy \cite{wang2024toothfairy} repurposed earphones to collect bone-conducted sound signals for detecting toothbrushing locations at a tooth-level resolution (90.5\%). By analyzing how the intensity of the vibration sound is attenuated after the propagation of different brushed teeth, the system achieved precise tracking but was limited to electric toothbrushes due to their consistent vibration patterns. Building on the similar principle of ToothFairy, SmarTeeth \cite{yang2025smarteeth} expanded the in-ear audio-based approach to manual toothbrushing. It used variations in the way the sound traveled through the skull and reached the ear canal based on the location of the brushing, allowing SmarTeeth to detect brushing activities beyond electric toothbrushes, making it a more generalizable solution for everyday users.

\textbf{Remarks.} Building upon these foundational studies, future research can further enhance earable-based toothbrushing monitoring in several directions:
\begin{itemize}
    \item \textbf{Dental Condition Detection}. The sound characteristics during brushing change when encountering dental issues such as cavities and root canal damage. Advanced acoustic analysis can help identify these anomalies, enabling early detection and preventive care.
    \item \textbf{Brushing Technique Classification.} Different brushing styles, such as up-and-down, side-to-side, or circular motions, produce distinct acoustic and motion signatures. Future systems can classify these techniques and provide feedback to encourage proper brushing habits.
    \item \textbf{Multi-modal Fusion for Enhanced Accuracy.} Combining acoustic sensing with IMU data, barometric pressure, and other physiological signals can improve the robustness of brushing detection, making the system more adaptive to different users and brushing conditions.
    \item \textbf{Child-Friendly Interactive Feedback.} Young children often struggle with proper brushing habits. By integrating real-time audio feedback, gamification, and voice guidance, earable technology can make brushing more engaging and effective for children.
\end{itemize}
Overall, by leveraging the ubiquity of earphones and advancements in acoustic and motion sensing, these future directions can revolutionize oral health monitoring, making it more accessible, accurate, and engaging for users of all ages.

\begin{table}[t]
\caption{Summary of earable systems for movement and activity.}
\label{tab: table_activity}
\renewcommand{\arraystretch}{1.2}
\begin{tabular}{p{2.1cm}p{2cm}p{3.5cm}p{4.3cm}p{1.8cm}}
\hline
\textbf{System} & \textbf{Sensor} & \textbf{Application} & \textbf{Advancement} &\textbf{Performance} \\ \hline
\multicolumn{5}{c}{\textbf{Toothbrushing}} \\ \hline
BrushBuds  \cite{yang2024brushbuds} & IMU & Brushing region detection & Brushing recognition via head movements (manual toothbrush only) & 84.3\% acc for 6 regions \\ \hline
ToothFairy  \cite{wang2024toothfairy} & Internal mic &  Brushing teeth localization & Tooth-level recognition (electric toothbrush only) & 90.5\% Acc \\ \hline
SmarTeeth  \cite{yang2025smarteeth} & Internal mic & Brushing region and surface detection & 16-surface tracking for both manual and electric toothbrushes  & 92.7\% for six regions, 75.6\% for 16 surfaces \\ \hline
\multicolumn{5}{c}{\textbf{Human Activity Recognition}} \\ \hline
Hu \etal  \cite{hu2024detecting} & Internal mic & Foot strike pattern classification & Detects overstride, forefoot, heel strikes & 87.8\% Acc \\ \hline
WalkEar  \cite{stuchbury2025walkear} & IMU & Gait parameter monitoring & Holistic spatio-temporal and asymmetry analysis &  5.1\% MAPE on gaits, 2\% MAPE on kinetic parameters\\ \hline
GCCRR  \cite{xu2024gccrr} & IMU & Gait segmentation & Segments gait cycles from short sequences & 80\% Acc\\ \hline
Bian \etal  \cite{bian2024earable} & Electrostatic sensor & Step counting & Low-power step detection via body electric field & 96\% Acc\\ \hline
Sepanosian and Incel  \cite{sepanosian2024boxing} & IMU & Boxing gesture recognition & Real-time boxing gesture detection & 96\% Acc\\ \hline
Moschina \etal  \cite{moschina2023vertical} & IMU & Jump height estimation & Jump height estimation from ear accelerometer & 0.04m MAE \\ \hline
IMUPoser  \cite{mollyn2023imuposer} & IMU & Full-body pose estimation & Uses any subset of consumer IMUs & 12.1 cm Mean Per Joint Vertex Error  \\ \hline
MobilePoser  \cite{xu2024mobileposer} & IMU & Full-body pose + translation & Real-time hybrid model for phones and wearables & 10.6 cm Mean Per Joint Vertex Error \\ \hline
EarDA  \cite{lyu2024earda} & IMU & Human activity recognition & Domain adaptation from phone to earables & 89\% Acc \\ \hline
\end{tabular}
\end{table}

\subsection{Human Activity Recognition}

Human Activity Recognition (HAR) has emerged with broad applications in fitness tracking, healthcare monitoring, rehabilitation, gaming, and user-context adaptation \cite{mollyn2023imuposer}. Recent advances in earable devices have opened new avenues for non-intrusive and continuous activity sensing. Earables are uniquely positioned on the head, allowing them to unobtrusively capture subtle body and head movements, vibrations, and even bone-conducted signals. Their integration into daily life and compatibility with off-the-shelf hardware make earables a practical platform for on-the-go activity recognition.

Building on this potential, Hu \etal~\cite{hu2024detecting} proposed a system that uses the in-ear microphone to detect different foot strike patterns during running: overstride, forefoot strike, and heel strike, by analyzing bone-conducted vibrations, achieving 87.8\% accuracy. 
WalkEar~\cite{stuchbury2025walkear} proposed a holistic system for spatio-temporal, kinetic, and asymmetry gait parameter monitoring with earable IMUs. 
GCCRR~\cite{xu2024gccrr} proposes a method for segmenting gait cycles from short IMU sequences using ear-worn sensors to support home-based rehabilitation. 
Bian \etal~\cite{bian2024earable} designed a low-power step-counting solution utilizing the body area electric field acquired by a novel electrostatic sensing unit.
For head gesture recognition, Sepanosian and Incel~\cite{sepanosian2024boxing} introduced a real-time boxing gesture recognition system using IMU sensors on OpenEarable. Meanwhile, Moschina \etal\cite{moschina2023vertical} explored earable accelerometers for vertical jump testing, achieving a mean absolute error of only 0.04m in estimating jump height. For more comprehensive motion tracking, IMUPoser \cite{mollyn2023imuposer} estimates full-body pose using any available subset of consumer devices (earbuds, phones, watches). Building on this, MobilePoser~\cite{xu2024mobileposer} integrates pose estimation and global translation using a hybrid deep learning and physics-based pipeline. It achieves real-time performance on smartphones and demonstrates robustness with as few as one earable device. Moving beyond individual actions, EarDA~\cite{lyu2024earda} proposes an adversarial domain adaptation framework to improve HAR accuracy and data efficiency on earables by transferring knowledge from smartphone datasets. It combines Bi-LSTM feature extraction with filtering techniques to mitigate noise from unpredictable head motion, achieving nearly 89\% accuracy.

\textbf{Remarks.} These studies collectively highlight the growing trend of earable-based HAR. However, challenges remain in dealing with sensor variability, movement noise, and limited labeled data. Future research should focus on multi-modal fusion, domain adaptation across users and devices, and large-scale deployment studies. In particular, developing standardized benchmarks and exploring on-device learning approaches will be key to bringing earable activity recognition into real-world applications at scale.
\section{Interaction}
\label{sec: Interaction}

\begin{figure}[t]
    \includegraphics[width = 1\textwidth]{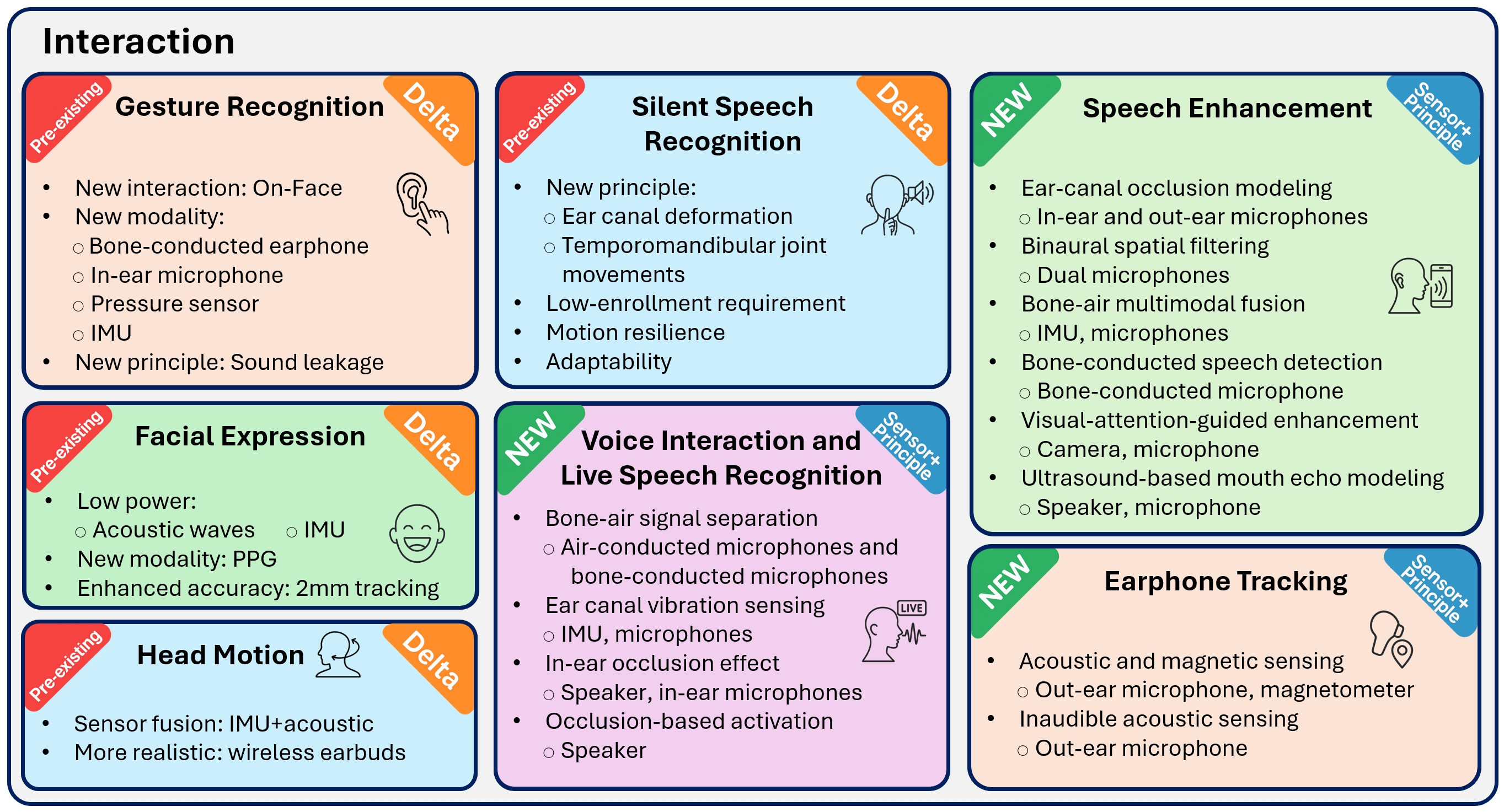}	
    \caption{Summary of recent advancements in
earable-based interaction.}
    \label{fig: Interaction}
    \Description{}
\end{figure}

This section summarizes recent advances in earable interaction modalities. As illustrated in Figure~\ref{fig: Interaction}, these works span both refinements of existing interaction paradigms, such as gesture recognition, facial expression, and head motion, and the emergence of new capabilities like speech enhancement, voice interaction, and earphone tracking. The evolution includes not only new sensing principles (e.g., sound leakage, TMJ motion, in-ear occlusion effects) but also novel sensor configurations that improve robustness, reduce power consumption, and enable more natural, context-aware interactions.

\subsection{Gesture Recognition}

Gesture recognition offers a more intuitive and socially acceptable alternative to traditional touch or voice interfaces. The underlying principle of gesture recognition in hearables involves interpreting physical interactions ranging from on-body touches to mid-air hand movements by leveraging sensors like IMUs, microphones, or pressure sensors embedded in or around the ear. Applications span from music control and call management to health tracking and context-aware computing, making gestures a compelling mode of interaction especially when hands are occupied.

Early work in this space largely relied on IMU-based gesture recognition, which remains a foundational direction. EarBender~\cite{alkiek2023earbender} leveraged commercial earables’ built-in IMUs to detect natural hand-to-ear gestures such as taps and swipes. Building on this, Sato \etal~\cite{sato2024exploring} emphasized user-driven personalization, conducting a Gesture Elicitation Study to collect user-defined gestures and designing an IMU-based recognition model that achieved over 91\% accuracy while accounting for personalized gestures and interaction-area constraints. Pushing further toward user-independent performance, Ui-Ear~\cite{zhao2024ui} introduced a novel on-face gesture recognition method based on vibration sensing. By combining IMU signals with domain-adversarial learning, they achieved strong cross-user generalization with 82.3\% accuracy.

To expand the sensing capability beyond IMUs, MAF~\cite{yang2024maf} employed active acoustic sensing using bone-conduction earphones, capturing perturbations in the emitted acoustic field caused by on- and near-face gestures. Compared to passive IMU methods, MAF supports a richer set of gestures with over 92\% recognition accuracy. Similarly exploring the acoustic domain, EarHover~\cite{suzuki2024earhover} proposed a mid-air gesture recognition method based on Doppler shifts in leaked audio. Unlike prior work focused on contact gestures, it achieved 88.8\% accuracy in real-world scenarios while remaining compatible with audio playback. 
 
For in-ear sensing, OESense~\cite{ma2021oesense} introduced a powerful method that exploits the occlusion effect inside the ear canal. By capturing low-frequency vibrations using inward-facing microphones, it robustly recognized subtle face tapping gestures with up to 97\% recall, even in noisy environments. Complementing this, Iguma \etal~\cite{iguma2023input} innovated with the use of atmospheric pressure changes within the ear canal, enabling detection of both touch and non-touch interactions (\eg, ear covering) with up to 99\% accuracy, offering a new, passive modality for robust interaction. From a design perspective, Panda \etal~\cite{panda2023beyond} expanded the interaction space by treating headphones as rich, sensor-enhanced surfaces. Their design exploration incorporated multi-modal sensing (capacitive touch, depth sensing, head orientation), showing how such integration can support complex gestures and contextual interactions, especially in scenarios like gaming and video conferencing.

\textbf{Remarks.} Gesture recognition for earables is rapidly evolving, driven by increasing hardware sophistication and creative sensing strategies. Research trends are shifting from simple IMU-based detection to robust user-independent recognition, exploring multimodal sensor fusion, and minimizing latency and energy consumption. Additionally, personalized gesture sets and on-device learning will play a critical role in ensuring high usability across diverse users and contexts. Finally, social acceptability and seamless integration with emerging XR and health applications remain crucial goals for widespread adoption.

\begin{table}[t]
\caption{Summary of earable systems for interaction.}
\label{tab: table_interaction}
\renewcommand{\arraystretch}{1.2}
\begin{tabular}{p{2.1cm}p{2cm}p{3.5cm}p{4.3cm}p{1.8cm}} 
\hline
\textbf{System} & \textbf{Sensor} & \textbf{Application} & \textbf{Advancement} &\textbf{Performance} \\ \hline
\multicolumn{5}{c}{\textbf{Gesture recognition}} \\ \hline
EarBender \cite{alkiek2023earbender} & IMU & Hand-to-ear gestures recognition & Little to no calibration & 97.4\% Acc \\ \hline
Sato~\etal \cite{sato2024exploring} & IMU & Gesture Personalization & User-driven customization & 91\% Acc \\ \hline
Ui-Ear \cite{zhao2024ui} & IMU & On-face interaction & User-independent & 82.3\% Acc\\ \hline
MAF \cite{yang2024maf} & Bone-conduction mic & On-face and near-face interaction & Using acoustic perturbations & 92\% Acc \\ \hline
EarHover \cite{suzuki2024earhover} & External mic & Mid-air gesture recognition & Compatible with audio playback & 88.8\% Acc \\ \hline
OESense \cite{ma2021oesense} & Internal mic & On-face interaction & Occlusion effect & 97\% recall \\\hline
Iguma~\etal \cite{iguma2023input} & Barometer & Covering the ear & Pressure change & 99\% Acc \\\hline
Panda~\etal \cite{panda2023beyond} & Touch, depth, IMU &  Context-aware gesture input & Rich design space exploration & N/A \\\hline
\multicolumn{5}{c}{\textbf{Head motion tracking}} \\ \hline
FaceOri \cite{wang2022faceori} & External-mic & Head pose estimation & Yaw/pitch sensing using inaudible chirps & 3.7° yaw, 5.8° pitch error \\ \hline
HeadTrack  \cite{hu2023headtrack} & External-mic & Real-time head tracking & Handles COTS earphone limitations (bandwidth, clock drift) & 4.9° yaw, 6.3° pitch error \\ \hline
IA-Track \cite{hu2023combining} & IMU, Microphone & Head tracking & IMU + acoustic hybrid approach for drift correction &  2 mm\\ \hline
\end{tabular}
\end{table}

\subsection{Head Motion Tracking}

Head tracking enables intuitive and immersive human-computer interaction (HCI), particularly in applications such as virtual reality (VR), augmented reality (AR), and the Metaverse. Traditional head-tracking methods primarily fall into three categories: vision-based, wearable sensor-based, and wireless signal-based~\cite{hu2023combining}. Vision-based systems often utilize external cameras to track facial landmarks or gaze direction, offering high accuracy but suffering from privacy concerns and limited field of view. Wearable devices such as VR headsets provide precise pose tracking but are typically bulky and expensive. Wireless signal-based approaches, including those leveraging Wi-Fi or ultrasound, offer more unobtrusive alternatives but are sensitive to environmental noise and often require dedicated infrastructure.

To overcome these limitations, recent work has explored earables as a platform for head tracking, capitalizing on their natural head placement and increasing market penetration of commercial off-the-shelf (COTS) earphones. Among these, IMU-based tracking is the most straightforward and cost-effective option but suffers from cumulative drift over time.   To address this issue, FaceOri~\cite{wang2022faceori} leverages ultrasonic ranging to estimate head pose using commercial earphones. By emitting inaudible chirps from a smartphone and measuring time-of-flight to the microphones embedded in earphones, the system could estimate head yaw and pitch. 

Building on this idea, HeadTrack~\cite{hu2023headtrack} tackled practical challenges of wireless earbuds, such as limited bandwidth and asynchronous clocks, and enhanced robustness through signal processing innovations. It achieved angular errors as low as $4.9^\circ$ (yaw) and $6.3^\circ$ (pitch) in real-time, enabling smart applications like gesture-based device switching and screen activation based on head orientation. Pushing further, IA-Track~\cite{hu2023combining} proposed a hybrid sensing approach by fusing IMU-based motion tracking with acoustic-based calibration. This technique corrected accumulated IMU drift via inaudible smartphone-emitted tones, combining the responsiveness of IMUs with the stability of acoustic ranging, and showcasing improved long-term tracking reliability. All these methods operate without external infrastructure, exploiting the spatial coupling between smartphones and earphones in everyday use.

\textbf{Remarks.} Earphone-based head tracking offers a compelling alternative to conventional systems, emphasizing usability, affordability, and infrastructure-free deployment. Compared to traditional vision or headset-based methods, it delivers a lightweight and scalable solution suitable for daily use. However, key challenges remain, such as handling dynamic acoustic noise, earphone misalignment, and generalization across users and devices. Future research should explore multimodal signal fusion (\eg, combining IMU, acoustic, barometric, and physiological signals like PPG) to boost robustness. Beyond HCI, potential applications in healthcare, cognitive state monitoring, and context-aware feedback could further elevate the value of this emerging sensing modality.

\subsection{Silent Speech Recognition}

Silent Speech Recognition (SSR) enables speech-based interaction without vocalization by detecting articulatory movements such as those from the lips, jaw, or tongue. Compared to conventional automatic speech recognition (ASR), SSR offers stronger noise robustness, better privacy, and accessibility for users with speech impairments. Early SSR efforts \cite{sahni2014tongue, chen2020c, khanna2021jawsense} using ear-mounted sensors, such as accelerometers, proximity sensors, or cameras, demonstrated the feasibility of recognizing silent speech via ear canal, cheek, or tongue-related deformations. However, these systems often required intrusive hardware (e.g., magnets \cite{sahni2014tongue} or cameras \cite{chen2020c}), relied on user-specific calibration \cite{khanna2021jawsense}, and lacked scalability for everyday use.

To overcome these limitations, recent studies have turned to earables equipped with built-in microphones and inertial sensors, which are already present in many commercial devices. These works eliminate the need for additional sensors while offering a more practical and unobtrusive form factor. Building on the principle of ear canal deformation, newer systems further improve recognition accuracy \cite{jin2022earcommand}, reduce the burden of user enrollment \cite{sun2024earssr}, and expand vocabulary size \cite{dong2024rehearsse}. Meanwhile, the use of temporomandibular joint (TMJ) movement as a complementary modality has emerged as a promising direction, enabling continuous and vocabulary-independent silent speech decoding using only head-worn sensors.

\begin{table}[t]
\caption{Comparison of representative silent speech recognition systems.}
\label{tab:ssr}
\renewcommand{\arraystretch}{1.2}
\begin{tabular}{p{2cm}p{2.3cm}p{6.5cm}p{3.5cm}}
\hline
\textbf{System} & \textbf{Vocabulary Scope} & \textbf{Key Contribution} & \textbf{Evaluation Highlight} \\ \hline

EarCommand \cite{jin2022earcommand} & Predefined commands \& sentences & First to combine ultrasound-based ear canal sensing with IMU-triggered activation for low-power SSR & 10.02\% WER (32 commands), 12.33\% (25 sentences) \\ \hline

EarSSR \cite{sun2024earssr} & Incremental letter \& word learning & Supports vocabulary expansion via incremental learning without retraining the full model & 82\% (letters), 93\% (words/phrases) \\ \hline

ReHEarSSE \cite{dong2024rehearsse} & Open-set with unseen word support & Generalizes to novel words using a temporal convolutional network trained on diverse examples & 89.3\% accuracy on 100 unseen words \\ \hline

HPSpeech \cite{zhang2023hpspeech} & Predefined commands & Demonstrates acoustic TMJ-based SSR using off-the-shelf headphones without hardware modification & >90\% accuracy on 8 commands \\ \hline

MuteIt \cite{srivastava2022muteit} & Large-vocabulary via phoneme recomposition & Uses a twin-IMU setup to infer phoneme sequences, enabling reconstruction of unseen words & 94.8\% accuracy on 100 words; robust to walking \\ \hline

Unvoiced \cite{srivastava2024unvoiced} & Free-form & Transforms IMU-based jaw motion to mel-spectrograms and decodes using an LLM, aligning SSR with ASR workflows & >94\% task completion; <9\% WER \\ \hline

Magnetic Skin \cite{dong2023decoding} & Phoneme-level & Achieves high-accuracy jaw tracking using a magnetometer and soft magnetic skin, avoiding face-mounted sensors & 93.2\% phoneme accuracy \\ \hline

QuietSync \cite{srivastava2024whispering} & Multilingual with few-shot tuning & Combines IMU and ExG signals to support language- and command-independent SSR with minimal calibration & Requires only 5 samples for adaptation \\ \hline

\end{tabular}
\end{table}

\textbf{Ear Canal Deformation-based.}
Ear canal deformation has been extensively studied for SSR, as subtle geometric changes in the ear canal correspond to different phonemes and words. Several recent works \cite{jin2022earcommand,sun2024earssr,dong2024rehearsse} adopt a common sensing principle: they emit inaudible ultrasound signals from the earphones, which reflect off the ear canal and are analyzed to extract deformation patterns associated with silent articulation. Building on this foundation, EarCommand \cite{jin2022earcommand} incorporates an additional IMU to detect user activity and trigger the SSR system, achieving a word error rate of 10.02\% for 32 word-level commands and 12.33\% for 25 sentence-level commands (N=12). However, it supports only a limited vocabulary and requires new users to provide a few enrollment samples. EarSSR \cite{sun2024earssr} improves flexibility by introducing incremental learning, enabling the model to adapt to new words while preserving previously learned knowledge, achieving 82\% accuracy on alphabet letters and 93\% on words and phrases (N=50). ReHEarSSE \cite{dong2024rehearsse} further advances generalization by recognizing previously unseen words using a modified temporal convolutional network, reaching 89.3\% accuracy on a 100-word vocabulary. While these systems demonstrate high accuracy and adaptability, their reliance on continuous ultrasound emission raises practical concerns. Specifically, always-on ultrasound may interfere with music playback, consume excessive battery power, and raise potential health concerns due to prolonged exposure to high-frequency sound. 

\textbf{Temporomandibular Joint Movement-based.}
Compared to ear canal deformation, TMJ movement offers a more direct physiological link to speech articulation, as it reflects jaw motion patterns that correspond closely to phoneme and word formation. As a result, recent works have increasingly focused on TMJ-based SSR, leveraging motion sensors or acoustic sensing to track subtle jaw dynamics. HPSpeech \cite{zhang2023hpspeech} uses the built-in speaker and microphone of commodity over-ear headphones to emit inaudible signals and detect TMJ movement, achieving over 90\% accuracy for eight silent speech commands (N=18). However, it still relies on active acoustic probing, which may raise similar concerns as ultrasound-based methods, such as interference with audio playback and energy consumption. To overcome these limitations, MuteIt \cite{srivastava2022muteit} introduces a purely IMU-based approach that avoids active sound emission altogether. It employs a twin-IMU configuration to isolate jaw articulation from head motion, enabling dynamic reconstruction of spoken content at the phoneme level rather than relying on fixed vocabularies. This improves vocabulary scalability and supports robust performance even during movement, achieving 94.8\% accuracy across 100 command words (N=20). Building on this, Unvoiced \cite{srivastava2024unvoiced} transforms IMU-detected jaw motions into mel spectrograms and applies a large language model (LLM) for text inference. This cross-modal design not only integrates seamlessly with existing ASR pipelines but also eliminates the need for custom classifiers, reaching over 94\% task completion with a word error rate below 9\% (N=19). These advances highlight the advantage of IMU-based TMJ sensing in enabling continuous, passive, and power-efficient SSR without compromising accuracy. Other methods have also explored hybrid sensing to enrich TMJ tracking. A soft magnetic skin-based system \cite{dong2023decoding} places a Bluetooth magnetometer behind the ear to capture jaw movements, achieving 93.2\% phoneme recognition (N=5) while avoiding face-mounted hardware. QuietSync \cite{srivastava2024whispering} combines IMUs with ExG electrodes on the earphones, glasses, and face to jointly sense jaw and muscle activity. It maintains robustness across languages, command lengths, and speaking styles, requiring only five samples for adaptation.

\textbf{Remarks.} Recent advances in SSR have focused on improving flexibility, generalization, and usability over earlier constrained systems.
\begin{itemize}
    \item \textbf{Phoneme-level Generalization:} Traditional SSR models often operated at the word level, requiring specific training data for each command. In contrast, newer systems like MuteIt \cite{srivastava2022muteit} and Unvoiced decompose silent speech into phoneme sequences, enabling the reconstruction of arbitrary words rather than classifying from a fixed list. This phoneme-based modeling significantly improves vocabulary generalization and supports dynamic command creation without the need for word-by-word retraining.
    \item \textbf{Lexicon Expansion:} Early SSR systems were typically constrained to a small, predefined set of command words, limiting their applicability in open-ended scenarios. Recent works such as EarSSR \cite{sun2024earssr}  and ReHEarSSE \cite{dong2024rehearsse} have addressed this limitation through incremental and open-set learning strategies. EarSSR allows continuous extension of the vocabulary without forgetting previously learned words, while ReHEarSSE further supports the recognition of previously unseen words by leveraging phoneme-level temporal patterns. These approaches mark a shift from closed-set classification to more scalable and flexible recognition models.
    \item \textbf{Modality Fusion and Pipeline Compatibility:} To improve robustness and expressiveness, several SSR systems have moved toward multi-sensor fusion and cross-modal integration. For example, Unvoiced converts IMU signals into mel spectrograms to interface directly with large language models, while QuietSync \cite{srivastava2024whispering} combines IMUs with ExG electrodes to capture both articulatory and muscular activity. These designs enhance signal richness and compatibility with standard ASR pipelines, paving the way for more seamless integration into existing voice-based systems.
    \item \textbf{Adaptability and Personalization:} Despite recent advances, SSR systems still require some degree of personalization, as articulatory signals, such as ear canal deformation or TMJ motion, exhibit strong individual variation. Some systems support lightweight fine-tuning (e.g., QuietSync \cite{srivastava2024whispering} requires only five samples), and others like EarSSR \cite{sun2024earssr} employ incremental adaptation. Nevertheless, reducing user calibration effort remains an important direction for improving usability.
\end{itemize}
Together, these developments mark a shift toward more scalable, adaptive, and integration-ready SSR solutions for real-world use.

\begin{table}[t]
\caption{Comparison of representative live speech recognition systems on earable platforms.}
\label{tab:liveSpeech}
\renewcommand{\arraystretch}{1.2}
\begin{tabular}{p{2cm}p{2.6cm}p{5.1cm}p{3.4cm}}
\hline
\textbf{System} & \textbf{Modality} & \textbf{Key Contribution} & \textbf{Evaluation Highlights} \\ \hline

VibLive \cite{li2024vibhead} & Bone-conducted microphones & Introduces continuous liveness verification by separating bone- and air-conducted signals, offering spoof-resistant voice authentication & 98.15\% liveness detection accuracy (N=27) \\ \hline

EarSpy \cite{cao2023live} & IMU & Uses in-ear IMU to detect ear canal vibrations caused by vocalization, enabling contact-based liveness verification under ambient noise & >87\% accuracy for live speech detection (N not reported) \\ \hline

EAROE \cite{han2025earoe} & In-ear microphone & Captures speech through in-ear body-channel sensing and reconstructs wideband audio using attention-based networks, achieving user-exclusive voice input & Outperforms baselines in intelligibility and fidelity (N=28) \\ \hline

EarVoice \cite{chen2024enabling} & Speaker (no mic) & Enables microphone-free voice interaction by detecting occlusion-induced echoes via earphone speaker, supporting speech input on ultra-low-cost devices & 89.5\% wake word accuracy; comparable to AirPods Pro \\ \hline

\end{tabular}
\end{table}

\subsection{Live Speech Recognition and Voice Interaction}

Voice-enabled interaction has become a defining feature of modern earables, supporting hands-free tasks such as voice assistant invocation, in-call commands, and passive listening. While traditional voice interfaces, most notably those on smartphones, have addressed concerns around spoofing attacks \cite{zhang2017hearing,zhang2020viblive}, they typically rely on air-conducted signals captured near the mouth and are rarely optimized for always-on, on-ear, or low-power operation. As earbuds increasingly become standalone platforms for speech input, recent systems revisit these security and privacy challenges through the lens of in-ear sensing, focusing on liveness verification, speaker isolation, and microphone-free interaction.

Some works focus on liveness detection, aiming to verify whether a speech signal is generated by a physically present user rather than a playback or spoofed source. VibLive \cite{li2024vibhead} introduces a continuous liveness verification mechanism using bone conduction signals. By simultaneously capturing air- and bone-conducted voice via built-in microphones and applying signal separation, the system isolates the bone-conducted component, which is tightly coupled to the speaker’s anatomy. Unlike conventional air-only approaches, this method captures body-coupled signals that are difficult to forge, effectively mitigating spoofing threats, and achieves a liveness detection accuracy of 98.15\% across 27 participants. Building on this concept, EarSpy \cite{cao2023live} leverages the earphone’s accelerometer to detect subtle vibrations and movements in the ear canal induced by live speech. By analyzing motion signals while filtering out head movement and speaker playback artifacts, EarSpy introduces a contact-based liveness modality that is robust to environmental audio, distinguishing live speech from external sound sources with over 87\% accuracy. This method demonstrates that even inertial sensors, traditionally used for motion tracking, can provide valuable liveness cues—but it also raises questions around motion-based side-channel leakage.

While the above methods focus on verifying speech authenticity, other research aims to ensure that only the wearer’s voice is captured and processed, thereby reducing the risk of false triggers and enhancing user privacy. EAROE \cite{han2025earoe} introduces a body-channel-based voice interface that captures speech through occlusion-induced in-ear vibrations, physically filtering out external voices while preserving internal speech structure. It employs an attention-based encoder-decoder network trained with physics-based data augmentation to reconstruct wideband speech from narrowband body-channel input. Compared to traditional microphones that indiscriminately capture ambient sounds, EAROE creates a user-exclusive speech input channel, and evaluated on 28 participants, it outperforms existing baselines in preserving both signal fidelity and intelligibility. Pushing the boundary toward minimal hardware deployment, EarVoice \cite{chen2024enabling} explores ultra-lightweight designs for hands-free voice activation under extreme constraints. Many low-cost earphones, such as those bundled with mobile devices, lack built-in microphones, rendering traditional voice interfaces infeasible. In contrast to these assumptions, EarVoice enables speech detection and user authentication using only the earphone’s speaker, leveraging occlusion-effect-induced acoustic patterns to detect speech activity and verify speaker identity through spectral cues. It further enhances wake word intelligibility by reconstructing high-frequency components from a stored template. Despite having no dedicated microphone, EarVoice achieves a wake word recognition accuracy of 89.5\%, rivaling commercial systems like AirPods Pro.

\textbf{Remarks.} Based on recent advances, we observe two notable paradigm shifts in how modern earable systems approach voice input, as well as several emerging opportunities for future development:
\begin{itemize}
    \item \textbf{From Air-Conducted Audio to Body-Coupled Signals.} Earlier voice interaction systems relied solely on air-conducted speech captured by microphones, making them highly susceptible to environmental noise, nearby speakers, and replay attacks. Recent systems, such as VibLive and EarSpy, shift toward body-coupled sensing (e.g., bone conduction and ear canal vibrations) to extract speech signals that are tightly bound to the physical presence of the speaker. This transition significantly improves liveness verification and environmental robustness, driven by the increasing need for secure voice interaction in mobile and shared spaces.
    \item \textbf{From Content-Only Capture to Speaker Isolation.} While conventional systems treat speech as purely acoustic content, newer designs (e.g., EAROE) treat the speaker’s body as a filter, capturing only the user’s speech through occlusion-induced in-ear vibrations while attenuating external sound. This physically grounded isolation mechanism ensures that only the wearer’s voice is captured, enabling private interaction even in noisy environments. This evolution is motivated by the need to prevent false wake-ups and protect verbal privacy during passive listening.
    \item \textbf{Future Directions.} Future systems should address user anatomical variability—e.g., differences in occlusion effect across ear shapes or bone density, which may affect signal consistency. They should also explore multimodal fusion, combining in-ear audio, motion (IMU), and biosignals (e.g., PPG or ExG) to improve robustness under motion and noise. For example, cross-domain liveness verification could fuse IMU and acoustic cues to reject both audio spoofing and silent impersonation; or zero-mic voice interfaces could combine speaker vibration and skin-conduction sensing to enable speech input on ultra-minimal earbuds.
\end{itemize}


\subsection{Speech Enhancement}
Speech enhancement has not been a major focus in earlier earable surveys \cite{roddiger2022sensing}, but recent hardware and usage trends have brought it to the forefront. As smartphone manufacturers phase out the 3.5mm headphone jack and users increasingly prioritize convenience, wireless earbuds have become the default choice for music playback, phone calls, and video conferencing. This shift in usage scenarios has raised new demands for high-quality uplink speech under real-world, noisy conditions. Compared to wired earphones with microphones located near the mouth, wireless earbuds face unique challenges: their microphones, commonly referred to as voice pickup units (VPUs), are positioned at the ear, farther from the speech source. As a result, the captured air-conducted speech signals suffer from attenuation and are more easily corrupted by environmental noise. At the same time, modern earbuds are equipped with multiple microphones (often used for active noise cancellation) and sensors such as IMUs and proximity detectors. This multi-modal hardware configuration creates new opportunities for advanced speech enhancement. The availability of both in-ear and out-ear microphones enables directional noise suppression, while auxiliary sensors can provide contextual information (e.g., motion, head orientation) to further inform signal processing. Together, these developments make earables a compelling and increasingly practical platform for robust speech enhancement.

\begin{table}[t]
\caption{Comparison of representative speech enhancement systems for earables.}
\label{tab:speech_enhancement}
\renewcommand{\arraystretch}{1.2}
\begin{tabular}{p{2cm}p{2.5cm}p{5.5cm}p{4.0cm}}
\hline
\textbf{System} & \textbf{Modality} & \textbf{Key Contribution} & \textbf{Evaluation Highlights} \\ \hline

ClearBuds \cite{chatterjee2022clearbuds} & Dual out-ear microphones & First to demonstrate synchronized binaural beamforming on wireless earbuds, enabling real-time spatial enhancement & Outperforms Conv-TasNet in user study (N=37); real-time on mobile \\ \hline

ClearSpeech \cite{ma2024clearspeech} & In-ear and out-ear microphones & Applies occlusion-aware deep fusion without hardware modification, improving speech quality under noise & Improved speech quality and intelligibility vs. single-mic baseline \\ \hline

EarSpeech \cite{han2024earspeech} & In-ear and out-ear microphones & Models occlusion-induced cross-channel correlation to achieve supervised-level enhancement using only 15 minutes of training data & Comparable to fully supervised systems with minimal training \\ \hline

In-Ear-Voice \cite{schilk2023ear} & Bone-conduction microphone & Demonstrates ultra-low-power VAD using bone-conducted vibrations instead of air-conducted signals & 95\% VAD accuracy; 12.8 ms latency; 14 µJ per inference \\ \hline

VibVoice \cite{he2023towards} & Microphone and IMU & Fuses air and bone-conducted signals with a synthesis pipeline to enable scalable multimodal training & 26\% SNR and 21\% PESQ gain; 31× faster than SOTA on mobile \\ \hline

EarSE \cite{duan2024earse} & Ultrasound & Captures facial articulation via ultrasound reflections as an auxiliary cue for speech enhancement & 38.0\% SI-SNR and 20.5\% PESQ gain; 22.45–66.41\% word error rate reduction \\ \hline

Look Once to Hear \cite{veluri2024look} & Microphone and visual attention & Leverages short-term gaze as a hands-free enrollment cue for target speaker extraction & 7.01 dB SI-SNRi gain with 1–4 sec enrollment; 18.2 ms latency on embedded CPU \\ \hline

\end{tabular}
\end{table}

\textbf{Microphone-based Approaches.}
Traditional speech enhancement methods often rely on multi-microphone beamforming to isolate the target speaker. Classical signal processing-based beamformers \cite{chhetri2018multichannel} are computationally lightweight and well-suited for embedded devices, but they struggle to capture the rich spectral-temporal dynamics of speech in real-world noise. On the other hand, deep neural networks \cite{luo2019conv} offer superior performance by learning complex patterns, yet their computational cost often exceeds what typical wireless earbuds can support. Recent works aim to resolve this trade-off by proposing efficient, learning-based enhancement models tailored to the constraints and configurations of earbud hardware.

ClearBuds \cite{chatterjee2022clearbuds} presents a model-driven enhancement system that utilizes a custom binaural earbud prototype capable of synchronized two-ear recording, forming a spatial microphone array. This cross-ear design enables head-worn beamforming, a spatial enhancement capability not supported by most existing commercial earbuds due to the lack of synchronized binaural recording. Compared to mono-channel learning models, ClearBuds leverages spatial diversity to better suppress background noise and improve speech intelligibility. It also adopts a hybrid training strategy combining synthetic mixtures, mannequin-based recordings, and real human speech, which enhances generalization across acoustic conditions. In real-time evaluations, ClearBuds achieves higher perceptual speech quality than a Conv-TasNet baseline while remaining lightweight enough to run on mobile processors.

In contrast to ClearBuds’ spatial design, ClearSpeech \cite{ma2024clearspeech} and EarSpeech \cite{han2024earspeech} explore a single-earbud architecture that leverages the complementary sensing capabilities of in-ear and out-ear microphones. Rather than performing stereo beamforming, both systems rely on cross-channel fusion: the out-ear microphone captures air-conducted speech exposed to environmental noise, while the in-ear microphone, benefiting from ear canal occlusion, provides a higher-SNR signal containing both air- and bone-conducted components. ClearSpeech enhances both magnitude and phase through a deep neural network, incorporating data synthesis and gating mechanisms tailored to the unique spectral characteristics of in-ear recordings. EarSpeech adopts a dual-branch architecture that explicitly models the occlusion-induced correlation between the two channels and employs a noise mixture training strategy to improve robustness in unseen conditions. Notably, both systems operate on off-the-shelf earbuds without requiring hardware modifications, offering a practical path toward commercial deployment. Despite differences in architectural choices, they collectively demonstrate the effectiveness of dual-microphone fusion in enhancing speech quality under challenging acoustic environments.

\textbf{Multi-modality Approaches.}
While microphone-based approaches remain central to earable speech enhancement, they often struggle under extreme noise or motion \cite{chatterjee2022clearbuds}, especially when relying solely on air-conducted signals. To address these limitations, recent studies have explored multimodal solutions that integrate bone-conducted vibrations, inertial motion, or visual cues. These additional sensing channels offer complementary information that enhances robustness, particularly in acoustically challenging or highly dynamic environments.

In-Ear-Voice \cite{schilk2023ear} focuses on ultra-low-power voice activity detection by embedding a bone-conduction microphone inside a custom in-ear device. Unlike traditional air microphones, this sensor directly captures mechanical vibrations generated by the user’s vocal tract, making it inherently resistant to ambient noise. The system uses a compact neural network to detect speech onset with 95\% accuracy, a latency of just 12.8 ms, and a per-inference energy cost of only 14 µJ. Compared to microphone-based VAD solutions, In-Ear-Voice demonstrates that bone-conducted sensing can offer high reliability with minimal energy consumption, making it well-suited for always-on scenarios.

VibVoice \cite{he2023towards} takes a fusion-based approach, combining air-conducted microphone signals with bone-conducted vibrations captured via an embedded accelerometer. Each modality has distinct advantages: while the microphone offers broadband information with noise, the vibration channel is noise-resilient but spectrally limited. To exploit this complementarity, VibVoice adopts a dual-branch neural architecture with modality-specific losses and introduces a physics-informed data augmentation technique to simulate bone-conducted input from existing audio datasets. This design yields significant performance gains—up to 26\% improvement in SNR and 21\% in PESQ—while achieving real-time performance on mobile devices. Compared to single-modality models, VibVoice better preserves speech quality under low-SNR conditions without sacrificing latency.

In addition, EarSE \cite{duan2024earse} explores a novel non-acoustic modality for speech enhancement by leveraging facial articulatory gestures. The system emits an ultrasound field from the earbud toward the user’s cheek. It receives reflected echoes, which are then decoded into low-dimensional articulatory features using a lightweight neural encoder. These features serve as an auxiliary stream to guide the speech enhancement model, particularly in severe noise conditions. Evaluated in real-world mobile scenarios, EarSE achieves a 38.0\% improvement in SI-SNR and a 20.5\% gain in PESQ, while reducing word error rate by 22.45–66.41\% across various noise conditions and user states. This demonstrates the viability of facial ultrasound sensing as a complementary signal for robust speech enhancement on earable devices.

Look Once to Hear \cite{veluri2024look} introduces an interaction-centric perspective by using short-term visual attention as an implicit enrollment cue for target speech extraction. The user glances at the speaker of interest, during which a brief speech segment is captured to generate a speaker embedding. This embedding guides a real-time extraction model that isolates the target voice regardless of gaze shifts or background interference. Unlike conventional keyword spotting or voice enrollment systems, this method enables hands-free, gaze-driven interaction and aligns with natural user behavior. By combining spatial cues from binaural microphones with efficient inference on embedded CPUs, the system achieves a 7.01 dB speech quality improvement with as little as one to four seconds of visual priming, and operates with 18.2 ms end-to-end latency.

\textbf{Remarks.} The systems discussed above reflect a clear departure from traditional single-microphone speech enhancement pipelines toward multimodal, hardware-conscious, and interaction-aware architectures uniquely suited for the constraints and opportunities of earables. To better characterize this progression, we identify four emerging technical dimensions and associated challenges:
\begin{itemize}
    \item \textbf{Hardware-Efficient Real-Time Deployment.} Wireless earbuds face strict limitations in computation, battery life, and thermal budget, especially for real-time applications like telephony or voice assistance. Consequently, there is a noticeable shift away from large, high-capacity models toward low-latency, energy-efficient architectures optimized for edge inference. ClearBuds \cite{chatterjee2022clearbuds} demonstrates that synchronized binaural beamforming—once thought too resource-intensive—can be achieved in real time on mobile hardware by leveraging tightly coupled spatial microphones and model-light processing. Look Once to Hear \cite{veluri2024look} pushes this further by delivering target speech extraction at just 18.2 ms end-to-end latency on embedded CPUs, while operating without external sensors or preprocessing. These examples illustrate how system design is increasingly driven not just by raw performance, but by real-world deployability on constrained devices.
    \item \textbf{Training Data Challenges.} Supervised learning-based speech enhancement relies heavily on large-scale, diverse, and precisely paired datasets. This becomes particularly challenging in earable contexts, where novel sensing modalities—such as in-ear microphones, bone-conduction sensors, and occlusion-dependent signals—do not have widely available datasets. To address this, recent systems explore modality-specific data augmentation and simulation techniques. ClearSpeech \cite{ma2024clearspeech} proposes an Out-to-In (O2I) model that learns to approximate in-ear signals from standard out-ear recordings, capturing occlusion-induced spectral transformations. VibVoice \cite{he2023towards} models the physical transfer from air-conducted sound to vibrational acceleration, enabling the generation of bone-conducted training signals from existing audio corpora. These strategies significantly reduce data collection burdens and allow scalable model training across users, form factors, and sensor configurations.
    \item \textbf{Modality Fusion for Robustness.} Microphone-only enhancement methods often degrade in environments with strong noise or reverberation. Recent systems increasingly fuse non-acoustic modalities—such as bone-conducted signals, motion sensing, or visual attention to more reliably isolate the user's voice. These modalities provide physically grounded features and help disambiguate target speakers in complex scenarios. Such fusion approaches represent a transition from purely acoustic enhancement to context-aware, multimodal speech reconstruction.
    \item \textbf{Toward Interaction-Aware Enhancement.} Beyond passive denoising, systems like Look Once to Hear \cite{veluri2024look} introduce interaction-aware enhancement, using gaze direction or short-term attention as a dynamic enrollment signal. This reflects a broader trend where speech enhancement becomes part of the user interface, adapting to conversational context and user intent. Future systems may further integrate user gestures, facial activity, or even physiological states to anticipate and enhance speech in an anticipatory fashion.
\end{itemize}

\subsection{Facial Expression}

Facial expressions serve as a rich medium for conveying emotion, intent, and user state, making them central to applications in affective computing, healthcare, and human-computer interaction~\cite{li2022eario}. While camera-based facial expression recognition is well-established, its dependence on lighting, privacy concerns, and limited mobility have motivated researchers to explore wearable alternatives~\cite{zhang2023earphone}. Recently, earables have emerged as a promising platform for detecting facial muscle movements through acoustic, inertial, and physiological signals captured in and around the ear. These approaches offer unobtrusive, low-power, and privacy-preserving solutions for continuous facial expression tracking.

Pioneering this direction, EarIO~\cite{li2022eario} introduced active acoustic sensing via frequency-modulated continuous wave (FMCW) signals to detect skin deformations caused by facial muscle movement. Compared to passive sensing, this approach enabled fine-grained expression reconstruction at low power consumption, establishing a new baseline for earable-based expression tracking. Building on this, IMUFace~\cite{yao2025imuface} showed that subtle ear motions resulting from facial expressions could be captured using embedded IMU sensors. Their system reconstructed 3D facial landmarks in real time with only 58 mW power usage, offering a passive and energy-efficient alternative to active acoustic sensing.

To further enhance spatial resolution and continuity, EARFace~\cite{zhang2023earphone} utilized in-ear ultrasonic reflections combined with a transformer model, achieving continuous facial landmark estimation with sub-2 mm accuracy, outperforming prior methods in both precision and wearability. This work also validated the comfort and feasibility of integrating such sensing into everyday earbuds. Exploring an alternative sensing modality, PPGface~\cite{choi2022ppgface} demonstrated that facial muscle activity modulates blood flow, detectable via in-ear photoplethysmographic (PPG) signals. Their system achieved emotion classification accuracy exceeding 93\%, even under real-world conditions, showcasing a novel physiological pathway for emotion recognition without relying on motion or acoustic data.

\textbf{Remarks.} The evolution of earable-based facial expression recognition reflects a shift from bulky, privacy-invasive systems to compact, multimodal, and real-world-ready solutions. Compared with earlier approaches, recent work has improved accuracy, energy efficiency, and continuous tracking capabilities. Looking forward, multimodal fusion, \ie, combining acoustic, IMU, PPG, and even barometric sensing, holds promise for improving robustness to sensor placement variability and individual differences. Key challenges ahead include low-latency on-device inference, modeling of subtle expressions, and adaptability to diverse anatomical structures. With continued integration into commercial earbuds, these systems may soon enable emotion-aware computing and personalized health monitoring in truly ubiquitous scenarios.

\begin{table}[t]
\caption{Summary of earable systems for facial expression and earphone tracking.}
\label{tab: table_facial}
\renewcommand{\arraystretch}{1.2}
\begin{tabular}{p{2.1cm}p{2cm}p{3.5cm}p{4.3cm}p{1.8cm}}
\hline
\textbf{System} & \textbf{Sensor} & \textbf{Application} & \textbf{Advancement} &\textbf{Performance} \\ \hline
\multicolumn{5}{c}{\textbf{Facial Expression}} \\ \hline
EarIO \cite{li2022eario} & External mic & Facial expression tracking & Tracks skin deformation via acoustic sensing & 25.9 MAE, 154 mW \\ \hline
IMUFace  \cite{yao2025imuface} & IMU & Facial landmark reconstruction & Real-time 3D landmark estimation via ear motion & 2.21 mm precision, 58 mW \\ \hline
EARFace  \cite{zhang2023earphone} & Internal mic & Facial landmark tracking & Transformer model on ultrasonic reflection & 1.83 mm error\\ \hline
PPGface  \cite{choi2022ppgface} & PPG & Facial expression recognition & PPG response to muscle movement for emotion classification & 93.5\% Acc \\ \hline
\multicolumn{5}{c}{\textbf{Wireless earphone tracking}} \\ \hline
MagSound  \cite{wang2023magsound} & External mic, Magnetometer & Earphone tracking & Combines acoustic ranging and magnetometer sensing to remove clock offset & mm-level 2D tracking \\ \hline
BLEAR  \cite{ge2024blear} & External mic & Earphone tracking and activity recognition & Frequency conversion to bypass BLE sampling limits & 3.37 cm error, 97\% Acc \\ \hline
\multicolumn{5}{c}{\textbf{Earable understanding}} \\ \hline
GrooveMeter  \cite{lee2023groovemeter} & Internal mic + IMU & Music engagement sensing & Detects reactions like humming, singing, nodding & 0.89 F1-score \\ \hline
Semantic Hearing  \cite{veluri2023semantic} & External mic & Sound source extraction & Spatially-aware target sound extraction across 20 classes & 6.56 ms latency \\ \hline
DeepEar  \cite{yang2022deepear} & Binaural mic (artificial ears) & Multi-source sound localization & Combines ITD and learned features with sector-based model & 93.3\% Acc\\ \hline
DeepBSL  \cite{el2023deepbsl} & In-ear mic & Personalized 3D localization & Learns user-specific HRTFs from mobile phone & $2.9^{\circ}$ azimuth, $1.4^{\circ}$ elevation error  \\ \hline
\end{tabular}
\end{table}

\subsection{Wireless Earphone Tracking}

Wireless earphones, originally designed for audio playback and communication, have evolved into promising platforms for motion tracking and context-aware interaction. Their compact form factor and close coupling with the body make them ideal for ubiquitous interaction. Compared with handheld smartphones or instrumented environments, earphones provide a more natural and user-friendly interface for tracking subtle hand gestures, handwriting, head movements, or spatial interactions.

However, tracking wireless earphones poses significant technical challenges. First, due to BLE (Bluetooth Low Energy) protocol constraints, many systems face limitations on sampling rate and audio compression, making it hard to adopt traditional acoustic sensing techniques~\cite{ge2024blear}. Second, the clock offset between the earphone and the host device introduces non-negligible drift, which severely degrades acoustic ranging performance over time. To address these limitations, MagSound~\cite{wang2023magsound} introduces a dual-modality approach that combines acoustic ranging with magnetic field sensing. The key insight is that commercial earphones contain strong built-in magnets for sound production. By measuring the magnetic field strength using a smartphone’s magnetometer, MagSound can predict the earphone's relative position free from clock offset. Experiments demonstrate that MagSound achieves millimeter-level 2D tracking accuracy and significantly improves handwriting recognition compared to acoustic-only baselines. BLEAR~\cite{ge2024blear} tackles a fundamental limitation of wireless earphone tracking: BLE's low sampling rate and lossy audio compression. To enable BLE earphones to receive ultrasonic signals ($\geq17$ kHz), BLEAR designs a novel frequency conversion scheme based on the nonlinearity of microphones. By mixing a high-frequency mask signal with the beacon signals, the system down-converts ultrasonic information into BLE-compatible bands. Evaluation with 8 users shows a mean tracking error of $3.37 cm$ and 97\% accuracy in recognizing 7 everyday user activities.

\textbf{Remarks.} Wireless earphone tracking presents a compelling frontier in ubiquitous sensing, transforming a daily accessory into a powerful spatial interface. Earlier works also explored using wired or customized earphones for sensing, but they typically assumed access to raw audio streams or bypassed BLE constraints ~\cite{wang2023magsound}. In contrast, recent systems explicitly embrace the limitations of commercial off-the-shelf (COTS) wireless earphones, pushing earphone tracking toward real-world deployment. Looking forward, we envision several directions for future work. First, cross-modal learning (\textit{e.g.}, combining magnetic, acoustic, IMU, and BLE RSS features) may further enhance robustness and generalizability. Second, low-power embedded implementations will be key to enabling continuous sensing on energy-constrained earable devices. Finally, future research should consider close collaboration with hardware manufacturers to optimize earphone designs for sensing, such as integrating higher-quality microphones, adopting new communication protocols, and employing compression algorithms that better preserve signal fidelity for acoustic sensing. As research continues to refine both the sensing models and the underlying hardware constraints, wireless earphones are poised to become not only listening devices but intelligent, motion-aware agents.

\subsection{Earable Understanding}

Earable devices offer a unique vantage point for capturing rich multimodal signals, especially acoustic cues from both the environment and the human body. Recent research has begun to explore how such devices can move beyond passive listening toward understanding, \ie, inferring user intent, behavioral states, or environmental context from audio signals. This shift toward "earable understanding" demands robust sensing, real-time signal interpretation, and localization under diverse and noisy real-world conditions.

GrooveMeter~\cite{lee2023groovemeter} proposed a system that detects vocal and motion reactions to music via earbuds. Using in-ear microphones and IMUs, it identifies behaviors such as humming, singing along, or head-nodding, demonstrating how earables can infer user engagement from subtle auditory cues. To enable earables to attend to meaningful sounds while suppressing distractions, Veluri~\etal~\cite{veluri2023semantic} proposed Semantic Hearing, which extracts target sound classes (\eg, sirens, alarms) from binaural input, preserving spatial cues. It supports 20 sound classes and generalizes well across users and acoustic scenes, enabling programmable auditory focus and situational awareness. DeepEar\cite{yang2022deepear} leverages binaural microphones with artificial ears to perform multi-source sound localization. Inspired by the human auditory system, it fuses handcrafted features (\eg, ITD) with latent representations via a multi-sector neural network. DeepEar can localize multiple concurrent sound sources and dynamically adapt to new acoustic environments using transfer learning. DeepBSL\cite{el2023deepbsl} achieves personalized 3D sound localization by collecting a small set of user-specific HRTFs using in-ear microphones and a mobile phone.

\textbf{Remarks.} These studies exemplify how earables are evolving from passive listening devices to intelligent agents capable of real-time perception and understanding. Future systems should integrate personalization pipelines to capture individual auditory characteristics (\eg, HRTFs, vocal styles) for holistic context awareness. Moreover, on-device deployment of deep audio models will be essential for privacy and latency. Enabling continuous, user-centered understanding from sound will be key to the next generation of earable intelligence.

\section{Authentication and Privacy}

\label{sec: Authentication}

\begin{figure}[]
    \includegraphics[width = 1\textwidth]{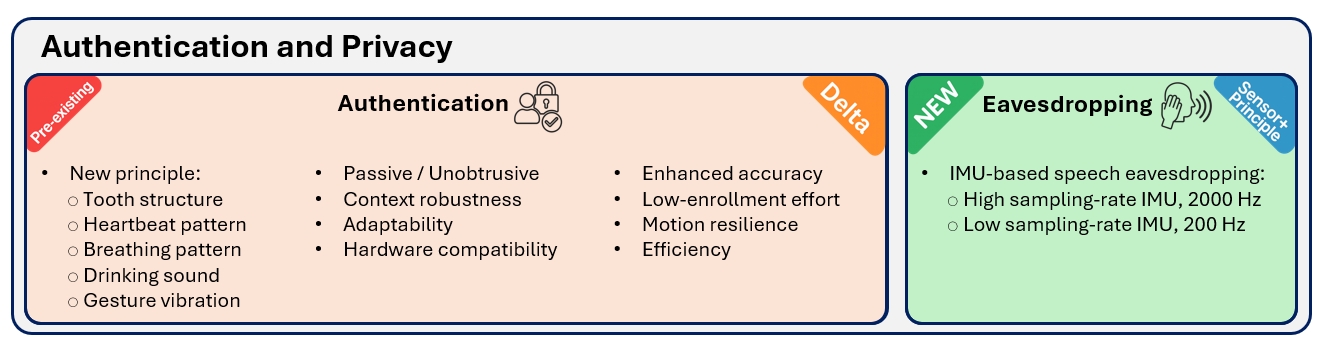}	
    \caption{Summary of recent advancements in
earable authentication and privacy.}
    \label{fig: Authentication}
    \Description{}
\end{figure}

This section presents recent advancements in earable authentication and privacy. As illustrated in Figure~\ref{fig: Authentication}, these developments span new biometric principles, practical improvements in system deployment, and emerging privacy threats such as IMU-based eavesdropping.

\subsection{Authentication}

With the growing adoption of earbuds as multifunctional devices, they are no longer used solely for audio playback. Modern earables increasingly process and store sensitive personal information, including voice commands \cite{feng2017continuous,ma2024clearspeech}, health data \cite{butkow2023heart,bui2019ebp,truong2022non}, and location history \cite{satongar2015influence}, while also serving as access points to online services, mobile devices, and IoT ecosystems. As such, securing access through user authentication has become an essential requirement for their safe and personalized use.

Previous earable authentication systems for earables primarily relied on biometric traits derived from the user’s anatomy and physiology, such as ear canal shape \cite{wang2021eardynamic,liu2014earprint,akkermans2005acoustic}, skull and jaw structure \cite{liu2021mandipass}, or bone-conducted vibrations \cite{ferlini2021eargate}. These systems typically used acoustic probing, body-transmitted signals, or inertial measurements to extract identity features, and achieved strong accuracy in controlled settings. However, these approaches also revealed critical limitations. Many required explicit user actions, such as speaking fixed phrases, remaining still, or covering the device with a hand. Others relied on audible or inaudible probe signals, which could interfere with audio playback or raise comfort concerns. Their performance was often sensitive to earbud placement, motion, and background noise, and most were not designed with real-time constraints, energy efficiency, or user experience in mind. As a result, they were difficult to apply in passive or continuous authentication scenarios during everyday use.

In response, recent works have proposed more passive, multimodal, and user-centric solutions. These systems leverage subtle and naturally occurring signals, such as in-ear audio transfer function \cite{hu2023lightweight,hu2024lr}, physiological rhythms like heartbeats \cite{cao2023heartprint,li2023earpass} or breathing \cite{han2023breathsign}, to enable seamless, low-effort authentication. Compared to earlier approaches, they place greater emphasis on hands-free and passive authentication, minimizing user burden during daily use. Many also draw inspiration from natural earable interaction behaviors, integrating signals from habitual gesture \cite{wang2024budsauth}, dental occlusion \cite{wang2022toothsonic,xie2022teethpass}, skin friction \cite{wang2024earslide}, and speech-induced variance \cite{duan2024f2key, choi2023earppg, srivastava2023jawthenticate, li2024piezobud}. Beyond accuracy, recent systems focus on real-world deployability, aiming to reduce enrollment effort \cite{hu2024lr}, latency \cite{srivastava2023jawthenticate, li2023earpass}, and power consumption while improving robustness to motion \cite{cao2023heartprint,han2023breathsign} and noise \cite{hu2023lightweight,hu2024lr,srivastava2023jawthenticate,wang2024budsauth}. Some approaches even support one-shot or continual learning \cite{cao2023heartprint}, enabling scalable deployment across users and environments. In the sections that follow, we categorize these modern approaches based on their core biometric principles, ranging from anatomical and physiological sensing to behavioral and voice-driven modalities, and discuss how they overcome prior limitations while advancing the usability and practicality of earable authentication.

\textbf{Ear Shape.}
The shape of the outer ear and ear canal has long been considered a natural biometric trait for user authentication, owing to its uniqueness and proximity to earable devices. Early approaches explored visual sensing by capturing the auricle’s geometry through external or front-facing cameras. Building on this idea, EarAuthCam \cite{mizuho2024earauthcam} integrates a miniature 3 mm² camera into the earphone to capture the ear contour for on-device authentication, achieving an EER of 8.36\% (N=20). While this design eliminates external hardware, it remains sensitive to lighting, hair occlusion, and device placement, which limits its robustness under real-world conditions, similar to earlier visual methods \cite{fahmi2012implicit,kohlakala2021ear}.

To overcome the visibility and occlusion issues of camera-based systems, other works have explored acoustic pathways. Amesaka \etal  \cite{amesaka2023user} repurpose the out-ear microphone to sense air leakage generated by audio playback, capturing variations influenced by ear canal and auricle geometry, as well as hand coverage. This method achieves 87.0\%–96.7\% BAC (N=16) by treating leakage patterns as biometric signatures. However, the system requires the user to press a hand over the earphone to enhance leakage—a step that significantly impairs usability and prevents passive operation.

Addressing both interference and interaction challenges, OnePiece \cite{gao2024eternity} proposes a fully hands-free approach. It still exploits air leakage but employs ultrasonic modulation schemes to reduce attenuation and multipath interference. This design achieves 94.6\% average accuracy (N=36) without requiring user actions, representing a clear usability improvement. Nonetheless, it continues to rely on actively modulated ultrasonic signals, which may interfere with ongoing audio playback or raise concerns in scenarios involving long-term continuous use. Compared to earlier visual and audible-leakage systems, OnePiece demonstrates notable progress in eliminating explicit user interaction and enhancing signal reliability through ultrasonic engineering. However, its reliance on active probing reflects a broader trade-off in ear shape–based methods between usability, audio compatibility, and sensing precision, highlighting the need for future designs that balance passive operation with minimal hardware intrusion.

\textbf{Ear Canal Geometry.}
The ear canal provides a personalized acoustic cavity that has long been explored for earable authentication. Earlier works primarily relied on actively emitted ultrasonic or audible probe signals to characterize this geometry \cite{gao2019earecho,mahto2018ear}. These signals, when reflected within the occluded ear canal, yield a transfer function that is distinctive across users. While such methods demonstrated strong discriminative power under clean conditions, they suffer from practical drawbacks: the active probing may interfere with music playback, reduce user comfort, and raise potential health concerns for prolonged use.

To address these limitations, Hu \etal  \cite{hu2023lightweight} proposed a fully passive approach that estimates the user’s unique occluded ear canal transfer function (OECTF) by comparing in-ear and out-ear recordings of naturally occurring audio—such as music or ambient speech. This design avoids injecting any stimulus signal and instead reuses existing audio content, making it non-intrusive and compatible with daily earbud use. It achieves an EER of 4.84\% even in noisy conditions (N=12), offering a practical path toward continuous, real-world deployment.

Building on this passive sensing foundation, LR-Auth \cite{hu2024lr} introduces multiple enhancements. First, it uses a dual-template strategy—separately optimized for noisy environments and music playback—to boost adaptability across use cases. Second, it adopts a lightweight cosine similarity matching technique, resulting in up to 400× reduction in latency and energy consumption compared to conventional neural models. With an EER below 0.2\% across five-frame input segments (N=30), LR-Auth balances accuracy, speed, and efficiency. Notably, it requires only a one-second composite sine wave stimulus during enrollment to establish a reliable user profile, dramatically simplifying the setup process.

While both Hu \etal  \cite{hu2023lightweight} and LR-Auth \cite{hu2024lr} exploit acoustic responses, EarCapAuth \cite{hanser2024earcapauth} takes a different route by sensing the canal’s electrostatic topology. Using 48 capacitive electrodes embedded in the earbud tips, it reconstructs a high-resolution 3D profile of the ear canal geometry, achieving an EER of 7.62\% (N=20). Although this method is passive and immune to acoustic interference, it requires significant hardware integration effort and additional power, limiting its feasibility in consumer-grade earbuds.

Compared to earlier ultrasonic probing methods, recent ear canal-based systems \cite{hu2023lightweight,hu2024lr}, show clear improvements in usability, playback compatibility, and system efficiency while preserving high recognition accuracy. By avoiding dedicated probe signals and leveraging naturally available audio or capacitive geometry, they represent a shift from lab-bound designs to practical, passive, and low-overhead authentication solutions. However, canal deformation due to mandibular motion during speaking or chewing remains a shared challenge, suggesting the need for motion-aware compensation in future designs.

\textbf{Skull and Teeth.}
Beyond the ear, a person’s skull and dental structure also serve as unique biometric identifiers. Prior studies have validated this principle using both air- and body-conducted signals during vocalization. For example, VoiceGesture \cite{gao2021voice} requires the user to speak, using the voice as a probing signal to excite the skull bone and tissue. It captures the resulting air- and bone-conducted signals from both ears for identity verification. Similarly, MandiPass \cite{liu2021mandipass} measures the vibrations propagating from the mandible to the ear using IMUs while the user vocalizes an “EMM” sound, capturing individual variations in skeletal transmission paths.

Building on this concept, HCR-Auth \cite{he2024hcr} introduces a new approach that directly uses a chirp signal (50–850 Hz) as an acoustic probe and employs bone conduction earphones along with an IMU positioned on the opposite side to capture the head contact response (HCR). This enables the system to extract biometric features related to the user's head tissue composition, including water, lipids, and fat-free solids. Notably, HCR-Auth leverages bone conduction earphones, which—unlike in-ear devices—lack a sealed ear canal and thus cannot benefit from the occlusion effect used in many prior earable authentication systems. By shifting to bone-conduction devices and designing for their acoustic limitations, HCR-Auth expands the design space for earable authentication. However, due to the IMU’s maximum sampling rate of 467 Hz, the probing signal remains in the audible range and may interfere with the listening experience. Furthermore, the system is not robust under dynamic postures such as walking or running, or when music is playing.

ToothSonic \cite{wang2022toothsonic} and TeethPass \cite{xie2022teethpass}, in contrast, authenticate users based on the unique sounds generated during dental occlusion. When a user bites down, occlusal sounds are absorbed, reflected, and dispersed by the skull before reaching the ear canals, where they exhibit individual differences due to variations in skull density and elasticity. Compared to “EMM”-based systems \cite{liu2021mandipass}, these methods are more discreet and socially acceptable, as they do not require audible vocalization and can blend into natural mouth movements.
ToothSonic introduces six sliding gestures and four tapping gestures to capture multi-level dental biometrics, achieving 92.9\% (N=25) accuracy. The design of diverse gestures enables greater flexibility and higher accuracy, especially as the number of enrolled users increases. However, the amount of unique biometric information varies across gestures, and requiring users to perform multiple actions may introduce cognitive or operational burden. TeethPass simplifies the process by allowing users to perform arbitrary occlusion actions without requiring specific gestures, though it does not support sliding. It achieves a BAC of 96.8\% (N=22), but its performance declines after eating or drinking, as food and beverages can adhere to the teeth and alter occlusal characteristics.

\textbf{Heartbeat and Breathing.}
Physiological signals—particularly heartbeat and respiration—offer a promising path for passive and unobtrusive authentication, as they are continuously generated by the body and inherently difficult to forge. Early work such as PPGPass \cite{cao2020ppgpass} demonstrated the feasibility of using wrist-based photoplethysmography (PPG) to capture pulse waveforms for user identification. However, wrist-worn sensors are susceptible to motion artifacts and skin contact quality, and integrating such sensing into earbuds remained technically challenging.

To address these issues, EarPass \cite{li2023earpass} relocates PPG sensing from the wrist to the in-ear canal, where the skin–sensor interface is more stable and better shielded from external light. It emits light into the ear canal and captures blood volume fluctuations through reflected optical signals, extracting features such as waveform morphology, inter-beat intervals, and frequency patterns. Compared to wrist-based PPG, this design offers improved wearability and signal consistency, enabling a seamless integration with existing earbud form factors. The system achieves 98.7\% authentication accuracy in static conditions (N=10), validating its feasibility for earable-based use.

Going beyond optical methods, HeartPrint \cite{cao2023heartprint} introduces a novel acoustic approach using in-ear microphones to capture low-frequency phonocardiograms (PCGs) transmitted through body tissues. This method benefits from the occlusion effect within the sealed ear canal, enhancing the signal-to-noise ratio of internal body sounds. Unlike PPG, microphone-based sensing is less affected by movement and does not require skin–sensor contact. Moreover, HeartPrint incorporates a continual learning mechanism that updates the user profile over time, supporting long-term use without re-enrollment. It maintains a false acceptance rate (FAR) of 1.6\% and false rejection rate (FRR) of 1.8\% across a 12-week period (N=45), demonstrating long-term adaptability and resilience to intra-user variation.

Breathing patterns have also been explored as biometric features, although they are weaker and more transient than cardiac signals. BreathSign \cite{han2023breathsign} leverages in-ear microphones to capture bone-conducted breathing acoustics, including normal, fast, and deep respiration. Despite the subtlety of these signals, the system achieves 95.17\% authentication accuracy using just one breathing cycle, with higher accuracy achieved when multiple cycles are aggregated. Importantly, BreathSign also demonstrates strong spoofing resistance, with a detection rate of 98.25\% against replay attacks (N=20), suggesting its viability for secure biometric systems.

Compared to traditional biometric modalities, heartbeat and respiration-based systems offer three critical advantages:
(1) They are fully passive—users do not need to perform any deliberate actions, and the signals are naturally and continuously available; (2) They are inherently private and difficult to spoof, as neither heart sounds nor breathing acoustics are externally perceptible; (3) They are well-suited for continuous authentication and on-device learning, as the system can use ongoing physiological signals between interactions to refine and validate user identity without explicit re-enrollment.

\textbf{Behavioral-based.}
Beyond static anatomical or physiological traits, recent work has explored user behavior as a biometric signal. Actions such as swallowing, facial contact, or finger gestures naturally occur in daily routines and exhibit person-specific patterns. Compared to passive signals, behavioral cues offer greater flexibility and user agency, enabling authentication to be embedded within familiar, habitual interactions.

SipDeep \cite{ahmed2024sipdeep} leverages swallowing as a biometric by capturing bone-conducted acoustic signals during drinking, reflecting variations in bone density, pharyngeal structure, swallowing technique, and ear canal acoustics. As swallowing is involuntary and difficult to imitate, SipDeep provides strong spoofing resistance while enabling seamless authentication during natural drinking behavior. It achieves 96.2\% accuracy for sip preferences and 96.5\% for gulp preferences (N=15).

In contrast, EarSlide \cite{wang2024earslide} uses an inward-facing microphone to capture frictional sounds generated when a user slides a finger across the face. By leveraging the face–ear channel, it encodes distinctive biometric traits such as ridge-groove depth, skin elasticity, and sliding dynamics. A Siamese neural network compares the extracted acoustic fingerprints with stored templates, achieving a BAC of 98.37\% with just a single gesture (N=26).

BudsAuth \cite{wang2024budsauth} takes a motion-based approach, using IMU sensors in the earbuds to detect subtle skin vibrations caused by face-touch gestures such as tapping or swiping. By modeling gesture-specific vibration patterns based on tissue mechanics, it achieves an exceptionally low EER of 0.03\% with seven consecutive gestures.

Compared to traditional biometrics, these methods are more user-controllable than passive physiological signals and offer greater resistance to spoofing than voiceprints. However, they must carefully balance gesture complexity, user effort, and continuity: while actions like swallowing or facial touches are easy to integrate into daily use, they may not support background or continuous authentication unless combined with complementary modalities.

\textbf{Multi-modal Voiceprint}
Traditional voice-based authentication systems in earables are highly susceptible to replay attacks, voice synthesis, and impersonation, limiting their security in open or adversarial environments. To address these vulnerabilities, recent research has moved toward multi-modal designs that combine vocal signals with complementary biometric cues—often derived from physiological or mechanical correlates of speech—to verify both who is speaking and how the speech is physically produced.

F2Key \cite{duan2024f2key} creates a facial acoustic sensing field by emitting ultrasound from the headphone speaker and capturing reflections with a boom or modular microphone on the opposite side. These reflections encode micro-movements of the articulatory muscles and facial structure during speech, forming a unique channel impulse response that is mapped to the speech signal using a generative model. By linking the acoustic content to speaker-specific facial dynamics, F2Key achieves strong resistance to spoofing—blocking 99.9\% of replay, 96.4\% of mimicry, and 95.3\% of hybrid attacks.

EarPPG \cite{choi2023earppg} takes a microphone-free approach, integrating PPG sensors to detect blood vessel deformations in the ear canal induced by facial muscle contractions during speech. This physiological signal reflects speaker identity at the muscular–vascular interface, forming a voice-aligned biometric signature that is immune to traditional audio attacks. EarPPG achieves 94.84\% accuracy and remains robust to acoustic noise, visual occlusions, and environmental lighting variability.

Jawthenticate \cite{srivastava2023jawthenticate} uses dual IMU sensors within earphones to capture bone-conducted vibrations traveling through the jawbone and skull during vocalization. These signals reflect individual skeletal structure and are inherently difficult to forge, offering strong resistance to deepfake and replay attacks. The system achieves 97.6\% accuracy but requires one IMU to be placed over the temporomandibular joint, posing integration challenges for standard earbud designs.

PiezoBud \cite{li2024piezobud} pushes further by embedding miniature piezoelectric sensors into earbuds to detect speech-induced skin micro-vibrations, capturing non-audible murmurs at a 10 kHz sampling rate—far exceeding the granularity of conventional IMUs. These high-frequency vibrations encode fine-grained vocal dynamics, enabling accurate authentication even in noisy or adversarial audio conditions. Despite hardware integration challenges due to sensor size and adoption maturity, PiezoBud requires only 15 seconds of enrollment, and achieves an EER of 1.05\% with just 0.06s latency, offering a highly practical and efficient solution.

Overall, these systems illustrate a critical shift: from voice-alone authentication toward multi-modal verification that ties speech content to user-specific physical, physiological, or vibrational traits. By grounding voiceprints in how speech is produced, rather than only what is said, they offer significantly improved security, spoofing resilience, and modality diversity.

\begin{table}[]
\caption{Summary of earable authentication systems.}
\label{tab: authentication}
\renewcommand{\arraystretch}{1.2}
\begin{tabular}{p{2.5cm}p{2.23cm}p{2cm}p{2cm}p{4.5cm}}
\hline
\textbf{System} & \textbf{Biometric Trait} & \textbf{Sensor} & \textbf{Performance} & \textbf{Key Improvement} \\ \hline

EarAuthCam \cite{mizuho2024earauthcam} & Ear shape & Camera & 8.36\% (EER) & New modality (miniature embedded camera) \\ \hline
Amesaka \etal  \cite{amesaka2023user} & Ear shape & Microphone & 96.7\% (BAC) & Passive method using acoustic leakage \\ \hline
OnePiece \cite{gao2024eternity} & Ear shape & Microphone, Speaker & 94.6\% (ACC) & Hands-free and compatible with commercial earbuds \\ \hline
Hu \etal  \cite{hu2023lightweight} & Ear canal & Microphone & 4.84\% (EER) & Lightweight, passive acoustic response \\ \hline
LR-Auth \cite{hu2024lr} & Ear canal & Microphone, Speaker & 99.8\% (BAC) & Context-robust, fast enrollment, dual template \\ \hline
EarCapAuth \cite{hanser2024earcapauth} & Ear canal & Capacitive electrodes & 7.62\% (EER) & New modality with spatial canal mapping \\ \hline
HCR-Auth \cite{he2024hcr} & Skull response & Speaker, IMU & 96.55\% (BAC) & Bone-conduction friendly, low enrollment \\ \hline
ToothSonic \cite{wang2022toothsonic} & Teeth & Microphone & 92.9\% (ACC) & Discreet, multi-level occlusion patterns \\ \hline
TeethPass \cite{xie2022teethpass} & Teeth & Microphone & 96.8\% (BAC) & Free-form gesture, unobtrusive \\ \hline
EarPass \cite{li2023earpass} & Heartbeat & PPG sensor & 98.7\% (ACC) & Passive, migrated to in-ear platform \\ \hline
HeartPrint \cite{cao2023heartprint} & Heartbeat & Microphone & 1.6\% (FAR), 1.8\% (FRR) & Continual learning, motion-resilient \\ \hline
BreathSign \cite{han2023breathsign} & Breathing & Microphone & 95.22\% (ACC) & Novel signal, spoof-resistant \\ \hline
SipDeep \cite{ahmed2024sipdeep} & Swallowing & Microphone & 96.2\% (ACC) & Involuntary action, hard to spoof \\ \hline
EarSlide \cite{wang2024earslide} & Sliding sound & Microphone & 98.37\% (BAC) & High accuracy via acoustic fingerprint \\ \hline
BudsAuth \cite{wang2024budsauth} & Face gesture & IMU & 0.03\% (EER) & Efficient, low-overhead vibration sensing \\ \hline
F2Key \cite{duan2024f2key} & Facial reflection & Microphone, Speaker & 99.9\% (ACC) & Multi-attack spoofing defense via CIR \\ \hline
EarPPG \cite{choi2023earppg} & Facial muscle & PPG sensor & 94.84\% (ACC) & Motion-resilient, microphone-free \\ \hline
Jawthenticate \cite{srivastava2023jawthenticate} & Jaw vibration & IMU & 97.6\% (ACC) & Replay-robust bone-conducted speech sensing \\ \hline
PiezoBud \cite{li2024piezobud} & Skin vibration & Piezoelectric sensor & 1.05\% (EER) & High-resolution sensing, 0.06s latency \\ \hline

\end{tabular}
\end{table}

\textbf{Remarks.}  
Building on earlier efforts that demonstrated the feasibility of biometric authentication via ear anatomy or body-conducted signals, recent systems have made significant strides in addressing the practical limitations of those early approaches. The progress can be understood along the following dimensions:

\begin{itemize}
    \item \textbf{Model Complexity.} Early systems often used handcrafted features or shallow classifiers, which limited performance in noisy or unconstrained settings. Recent approaches improve accuracy using deep neural networks (DNNs), but this comes at the cost of increased latency and energy consumption, particularly problematic for always-on, resource-constrained earables. Newer works such as LR-Auth \cite{hu2024lr} and PiezoBud \cite{li2024piezobud} have begun addressing this by employing lightweight architectures and efficient inference pipelines, paving the way for real-time, on-device authentication.
    \item \textbf{Enrollment Effort.} A practical challenge is the data requirement for profile creation. Many authentication systems require users to provide a substantial amount of enrollment data to build a reliable identity profile,such as performing repetitive gestures \cite{wang2024budsauth} or remaining still for extended periods \cite{cao2023heartprint}, which can be inconvenient and may discourage adoption. Future authentication frameworks should aim to reduce enrollment effort, possibly by incorporating few-shot learning, self-adaptive models, or transfer learning to enable accurate authentication with minimal user input.
    \item \textbf{Noise-resilient Concern.} While many systems perform well in controlled settings, real-world robustness remains a concern. Microphone-based methods are sensitive to ambient noise, and IMU/PPG-based systems may degrade during motion. Recent works address this in two ways: (1) Systems like HeartPrint \cite{cao2023heartprint}, and BreathSign \cite{han2023breathsign} leverage the occlusion effect to enhance internal signals and suppress noise; (2) Other systems \cite{hu2023lightweight, hu2024lr}creatively use environmental noise as a passive probing signal, enabling context-aware authentication, though performance depends on environmental sound richness and volume.
    \item \textbf{Continual Learning.} Earlier systems often treated biometric traits as static, assuming that identity features would remain consistent across time. However, real-world deployment reveals that both physiological and behavioral patterns, such as heartbeat rhythms, respiratory patterns, or gesture dynamics, can drift due to changes in health, mood, or usage conditions. To address this, recent systems like HeartPrint \cite{cao2023heartprint} introduce continual learning mechanisms that incrementally update biometric templates, allowing models to adapt without explicit re-enrollment. Notably, modalities such as heartbeats and breathing offer continuous, passive streams of data during wear, enabling ongoing refinement of the user profile through unobtrusive and natural interactions. This makes them particularly well-suited for adaptive, long-term authentication, supporting both robustness and user convenience.
    \item \textbf{Active and Passive Mechanisms.} Authentication methods can be broadly categorized as passive or active. Passive methods rely on biometric traits such as ear shape, heartbeat, or breathing, enabling continuous and implicit verification. Active methods, in contrast, are based on deliberate actions like gestures, teeth tapping, or speech, offering intuitive on-demand authentication. A promising direction is to design hybrid systems that dynamically switch between modes based on context, using passive sensing during normal use and active input in high-security scenarios.
\end{itemize}
In sum, recent works have made notable progress toward making earable authentication more practical and user-friendly. Addressing computational cost, enrollment burden, environmental robustness, and long-term adaptability will be key to realizing reliable, real-world deployments. Future systems should also strive for flexible modality design that adapts to user context seamlessly.

\subsection{Earphone Eavesdropping}

Modern earphones equipped with motion sensors can capture subtle vibrations caused by speaking, including bone conduction vibrations (BCVs) and ear canal dynamic motions (ECDMs)~\cite{cao2023can}. These signals can leak sensitive information such as spoken content, speaker identity, and gender without using a microphone~\cite{gao2023practical}. Because motion sensors often don’t require explicit permissions, they pose serious privacy risks as side channels.

EarSpy~\cite{cao2023can} leverages high-frequency motion sensor data (up to 2000 Hz) to recognize live speech from the user. They separate speech signals from noise such as body movement and earphone playback, enabling accurate, user-independent speech recognition. Gao \etal~\cite{gao2023practical} explore a more realistic scenario with motion sensors limited to 200 Hz. They show that even with lower sampling rates, speech, identity, and gender can still be inferred using raw data and a channel attention mechanism. Their model performs well across environments and speaking volumes, highlighting the persistent risk even under system-imposed constraints.

\textbf{Remarks.} As motion sensors become increasingly sensitive and ubiquitous in consumer devices, the risk of side-channel eavesdropping demands urgent attention from both researchers and manufacturers. Future research should focus on developing practical defenses against motion sensor-based eavesdropping. This includes designing OS-level monitoring to detect abnormal sensor usage and exploring learning-based countermeasures that can detect or distort potential leakage. At the same time, it’s important to balance privacy with legitimate sensing applications, such as health tracking or accessibility features, to ensure innovation continues safely.

\begin{table}[t]
\caption{Summary of earable systems for Eavesdropping.}
\label{tab: table_eavesdropping}
\renewcommand{\arraystretch}{1.2}
\begin{tabular}{p{2.1cm}p{2cm}p{3.5cm}p{4.3cm}p{1.8cm}}
\hline
\textbf{System} & \textbf{Sensor} & \textbf{Application} & \textbf{Advancement} &\textbf{Performance} \\ \hline
EarSpy  \cite{cao2023can} & IMU (up to 2000 Hz) & Live speech recognition & Separates speech from noise; user-independent model & 85\% precision \\ \hline
Gao \etal  \cite{gao2023practical} & IMU (200 Hz) & Speech, identity, gender inference & Channel attention on raw low-rate IMU; robust in the wild & 80.3\% precision \\  \hline
\end{tabular}
\end{table}
\section{Resources for Future Earable Research}
\label{sec: Resources}

Recent advances in earable sensing have sparked the development of a diverse ecosystem of supporting resources that enable researchers to prototype new ideas, train machine learning models, and ensure real-world robustness. This section reviews key enablers for future research in this space, including hardware platforms, curated datasets, and signal quality assessment tools. These resources form the foundation for scalable, reproducible innovation in ear-centric sensing and interaction.

\subsection{Hardware Platforms}
\label{subsec:hardware-platforms}

\begin{table}[t]
\caption{Summary of recent research-driven earable hardware platforms.}
\label{tab: table_earable_hardware}
\renewcommand{\arraystretch}{1.2}
\begin{tabularx}{\linewidth}{p{2.2cm}p{1.6cm}p{3.5cm}p{1.5cm}p{1.5cm}p{3.5cm}}
\toprule
\textbf{Platform (Year)} & \textbf{Trans- mission} & \textbf{Sensors} & \textbf{Memory} & \textbf{Battery Life} & \textbf{Special Features} \\
\midrule
\textbf{eSense} (2018) \cite{kawsar2018earables} & Bluetooth Classic & 6-axis IMU (1), Microphone (1) & None & 1.2 h & Real-time multimodal streaming, compact form factor \\
\addlinespace
\textbf{ClearBuds} (2022) \cite{chatterjee2022clearbuds} & Custom BLE & Microphone & 1 GB & 40 h & Stereo synchronized recording, sub-64 $\mu$s sync error \\
\addlinespace
\textbf{EarAce}  (2023) \cite{cao2023earace} & Wi-Fi, Bluetooth & Stereo Microphones (2) & microSD & Varies & ANC control, acoustic impedance profiling, motion artifact reduction \\
\addlinespace
\textbf{OpenEarable 1.3}  (2022) \cite{roddiger2022openearable} & Bluetooth BLE & 6-axis IMU (1), In-ear Ultrasonic Mic (1), Pressure Sensor (1), Temp Sensor (1) & microSD & 10 h & Modular design, ultrasonic ear canal profiling \\
\addlinespace
\textbf{OpenEarable 2.0}  (2025) \cite{roddiger2025openearable} & Bluetooth BLE Audio & 9-axis IMU (1), 3-axis Ear Canal IMU (1), Microphones (3), PPG (1), Temp Sensor (1), Pressure Sensor (1) & microSD & 8 h & Multi-sensor integration, BLE audio, two ExG variants available, on-device ML \\
\addlinespace
\textbf{OmniBuds}  (2024) \cite{montanari2024omnibuds} & Bluetooth BLE Audio & 9-axis IMU (1), PPG (1), Temp Sensor (1), Microphones (3) & 1 GB Flash + 8 MB RAM & 8 h & Cuffless blood pressure estimation, on-device ML (CNN accelerator) \\
\bottomrule
\end{tabularx}
\end{table}


In recent years, a growing number of research-driven earable platforms have emerged, offering distinct combinations of sensing modalities, system openness, and extensibility. Most of these platforms have been introduced after 2022, reflecting a recent surge of interest and advancement in earable hardware (with eSense being an earlier effort from 2018). While prior survey \cite{roddiger2022sensing}, has extensively summarized application trends, they have not systematically covered the evolution of hardware platforms. To address this gap, Table~\ref{tab: table_earable_hardware} summarizes several representative hardware designs developed by the academic community. To complement the table, we further describe below the key sensing capabilities and system features of selected representative platforms.

\begin{itemize}
    \item \textbf{eSense.} eSense \cite{kawsar2018earables} is one of the earliest open research platforms to demonstrate real-time, multimodal sensing using in-ear form factors. It integrates a 6-axis inertial measurement unit (IMU), an in-ear microphone, and dual-mode Bluetooth communication within a compact earbud-like enclosure. The platform enables real-time streaming of motion, acoustic, and BLE beacon signals via open APIs, supporting applications such as head gesture recognition, activity monitoring, and proximity-based interaction. eSense captures IMU data at 20 Hz and audio at 16 kHz, and includes a small rechargeable battery that supports up to 1.2 hours of continuous operation. 
    \item \textbf{ClearBuds.} ClearBuds \cite{chatterjee2022clearbuds} integrates custom wireless earbuds designed to function as a synchronized binaural microphone array, enabling stereo voice capture with sub-64 ${\mu}$s synchronization error between channels. Each earbud houses a PDM microphone, a BLE SoC, and runs a custom low-latency transmission protocol that streams dual-microphone data to a mobile device. The hardware supports 15.625 kHz sampling and operates up to 40 hours on a coin cell battery. ClearBuds is among the first systems to overcome the limitation of single-channel Bluetooth audio in commercial earbuds, enabling real-time, synchronized dual-channel data collection for mobile speech enhancement.
    \item \textbf{EarAce.} EarAce \cite{cao2023earace} is a compact plug-in platform designed to work with commercial active noise cancellation (ANC) earphones, enabling versatile acoustic sensing without needing self-built devices. It integrates a high-resolution stereo codec (ES8388), supports microphone gain tuning, ANC control, and dual-ear sensing. A key hardware innovation is its support for precise device wearing state profiling using acoustic impedance measurements and its motion interference elimination algorithm leveraging dual-ear asymmetry. It also features local SD card storage, wireless data streaming via Wi-Fi/Bluetooth, and a control app for flexible configuration.
    \item \textbf{OpenEarable 1.3.} OpenEarable 1.3 \cite{roddiger2022openearable} is a fully open-source, Arduino-compatible research platform featuring a 6-axis IMU, an inward-facing ultrasonic microphone, an in-ear pressure and temperature sensor, speaker, LED, and push button. It is designed with a low-cost, modular architecture using off-the-shelf components and 3D-printed parts, supporting up to 10 hours of BLE-based data streaming. Its ear-hook PCB form factor ensures stable wear and accurate pressure capture for applications like motion tracking, jaw movement sensing, and ultrasonic ear canal profiling.
    \item \textbf{OpenEarable 2.0.} OpenEarable 2.0 \cite{roddiger2025openearable} is a fully open-source, sensor-rich earable platform designed for physiological and behavioral sensing. Compared to version 1.3, it introduces major upgrades including three microphones (in-ear, out-ear, and bone-conduction), a dual inertial system with both a 9-axis IMU and a 3-axis ear canal accelerometer, and additional sensors such as a pulse oximeter, optical temperature sensor, and ear canal pressure sensor. It supports BLE audio, has onboard storage via microSD, and is extensible through 14-pin and 12-pin connectors. It also integrates an neural network accelerator (Tensilica HiFi 3z DSP Core) for on device deep learning inference. The authors also introduced two specialized variants equipped with electrodes \cite{knierim2023openbci} and biopotential sensing \cite{lepold2024openearable} capabilities for EEG, EMG, and EOG; more details are available in the original publication.
    \item \textbf{OmniBuds.} OmniBuds \cite{montanari2024omnibuds} feature a symmetric dual-ear design with a 9-axis IMU, a three-wavelength PPG sensor (green, red, IR), a medical-grade infrared temperature sensor, and three microphones (2 outer, 1 in-ear), all managed by an RTOS-based software stack. The platform supports on-device inference via a CNN accelerator and BioHub co-processor, offers 1GB flash and 8MB RAM per earbud, and achieves up to 8 hours of continuous sensing. Unique features include cuffless blood pressure estimation and multi-modal sensing with onboard machine learning, making OmniBuds a highly integrated and programmable research platform.
\end{itemize}
Most provide access to raw sensor data, and custom APIs.
This openness allows researchers to prototype new sensing algorithms, evaluate cross-modal interactions, and reconfigure system pipelines. Across the listed platforms, we observe growing support for multimodal integration (e.g., microphones, IMUs, PPG, and temperature sensors), on-device storage, and real-time programmability, reflecting the increasing maturity of earable research infrastructure and its suitability for diverse applications ranging from health monitoring to interaction sensing.

\subsection{Datasets}

To support algorithmic or deep learning based earable research, datasets provide essential resources for training models, validating techniques, and ensuring reproducibility. However, unlike vision or speech data, earable signals are highly user- and context-dependent, influenced by anatomy, fit, motion, and hardware variability. These factors generalize and reuse across studies particularly challenging.

To help address this, several curated datasets have been released in recent years. Table~\ref{tab: Dataset} summarizes six representative datasets, highlighting their sensing modalities, subject coverage, and application focus. We describe each dataset in detail below, focusing on its design, scope, and research utility.

\begin{table}[t]
\caption{Summary of representative earable datasets.}
\label{tab: Dataset}
\renewcommand{\arraystretch}{1.2}
\begin{tabular}{p{2.2cm}p{2.8cm}p{1.2cm}p{3cm}p{5cm}}
\toprule
\textbf{Dataset} & \textbf{Modality} & \textbf{Subjects} & \textbf{Duration / Samples} & \textbf{Application Scenarios} \\
\midrule
FatigueSet~\cite{kalanadhabhatta2021fatigueset} & EEG, PPG, IMU, video & 29 & 2 × 40 min sessions & Mental fatigue monitoring (MSIT task), fatigue labeled by KSS and NASA-TLX \\
\addlinespace
EarSet~\cite{montanari2023earset} & PPG, IMU & 30 & 16 motion tasks per subject & Motion-induced PPG artifacts under facial, head, and body movements \\
\addlinespace
EarSAVAS~\cite{zhang2024earsavas} & In-ear and out-ear microphones, IMU & 42 & 44.5 hours & Subject-aware vocal activity (8 vocal actions, noisy and quiet conditions) \\
\addlinespace
OESense~\cite{ma2021oesense} & In-ear microphone, IMU & 31 & 52k steps, 465 min HAR, 20,880 gestures & Step counting, HAR (5 tasks), face gestures (12 locations × 60 reps) \\
\addlinespace
EarGate~\cite{ferlini2021eargate} & In-ear microphone & 31 & 52,046 steps (26,023 cycles) & Gait-based authentication under 8 walking conditions \\
\addlinespace
ClearSpeech~\cite{ma2024clearspeechDataset} & In-ear and out-ear microphones & 20 & 9.4 hours clean; synthetic noisy data & Dual-mic speech enhancement using paired in-ear and out-ear recordings \\
\bottomrule
\end{tabular}
\end{table}

\textbf{FatigueSet.}
FatigueSet \cite{kalanadhabhatta2021fatigueset} is a multi-modal dataset developed to study mental fatigue through in-ear and head-worn sensing. It includes data from 29 participants who completed two 40-minute sessions alternating between resting and cognitively demanding tasks based on the Multi-Source Interference Task (MSIT), alongside self-reported fatigue levels using NASA-TLX and the Karolinska Sleepiness Scale (KSS). The dataset comprises synchronized recordings from four modalities: EEG from a Muse S headband (forehead and ear region), in-ear PPG and IMU from a custom earpiece (with photodiodes, green LEDs, and a 6-axis sensor), facial videos for expression tracking, and reaction-time task logs. All sensor streams are precisely aligned to task events, supporting fine-grained analysis of how fatigue develops over time. Additionally, the authors release baseline models showing that in-ear PPG and IMU signals alone can yield meaningful fatigue predictions, underscoring the feasibility of passive mental state monitoring with earables.

\textbf{EarSet.}
EarSet \cite{montanari2023earset} is a publicly available, high-resolution multi-modal dataset designed for motion-robust in-ear physiological sensing. It focuses on the ear canal—a site known for higher signal fidelity than wrist-based PPG but also greater vulnerability to motion artifacts and wearing variation. The data was collected using a custom earbud prototype embedding a three-wavelength PPG sensor (green, red, infrared) and a 6-axis IMU, worn bilaterally in soft silicone tips. Recordings were gathered from 30 participants performing 16 structured motion tasks, spanning facial expressions, head movements, and full-body activities. All signals are time-aligned (100 Hz for PPG, 50 Hz for IMU), and each segment is labeled with task boundaries for fine-grained analysis. In addition to raw signals, the dataset includes motion scores and beat-to-beat heart rate estimates, supporting applications such as artifact detection, signal enhancement, and sensor fusion under realistic, dynamic conditions.

\textbf{EarSAVAS.} EarSAVAS \cite{zhang2024earsavas} is the first publicly available dataset explicitly designed for subject-aware vocal activity sensing on earables, addressing the lack of subject attribution and wearable-based data in prior datasets. It includes 44.5 hours of synchronized audio and IMU recordings from 42 participants in 23 dyads. Each dyad alternated between performing eight vocal activities, such as coughing, chewing, and speaking, while the partner remained silent, enabling clear subject labeling. Data were collected using modified commercial earbuds equipped with an inertial sensor and dual microphones, including an in-ear feedback mic that captures bone- and body-conducted signals unique to the wearer. Recordings were conducted under both quiet and noisy conditions, with ambient noise sourced from AudioSet and ESC-50. All signals are time-aligned and labeled with activity type and speaker identity. The dataset is accompanied by EarVAS, a lightweight multimodal model that fuses IMU and microphone data, achieving 90.84\% accuracy and 89.03\% Macro-AUC. EarSAVAS supports a wide range of applications in speech-driven interaction and privacy-aware earable sensing.

\textbf{OESense.} The dataset accompanying the OESense paper \cite{ma2021oesense} supports in-ear motion sensing using the occlusion effect. The authors modified a pair of wired commercial earbuds by embedding inward-facing MEMS microphones to capture bone-conducted sound within the occluded ear canal. Data were collected from 31 participants across three tasks: (1) step counting, with 52,047 annotated steps across varied walking conditions; (2) human activity recognition, covering five actions (walking, running, still, chewing, and drinking), yielding 465 minutes of labeled data; and (3) hand-to-face gesture recognition, involving 12 facial tapping gestures, repeated 60 times per participant, totaling 20,880 gestures. All signals were sampled at 48 kHz (microphone) and 100 Hz (accelerometer), and are annotated with ground truth activity labels. The dataset demonstrates the robustness of inward-facing microphones to motion and noise and serves as a resource for activity classification, gesture recognition, and motion-interference–resilient sensing. A Kaggle release \cite{ma2021oesenseDataset} provides access to raw data, annotations, and benchmark results.

\textbf{EarGate.} The dataset accompanying the EarGate study \cite{ferlini2021eargate} focuses on gait-based user identification using in-ear microphones that exploit the occlusion effect. The authors modified commercial wired earbuds by embedding inward-facing microphones near the ear tips to record bone-conducted acoustic signals during walking. Data were collected from 31 participants (15 male, 16 female) walking under eight combinations of ground type (tile or carpet) and gait condition (barefoot, sneakers, slippers, or walking while speaking). Each participant performed 1.5-minute walking sessions per condition, resulting in a total of 52,046 labeled steps (26,023 gait cycles). Microphone data were sampled at 48 kHz and annotated with walking condition and participant ID. The dataset enables evaluation of acoustic gait-based authentication models and includes baseline results using SVM classifiers. It is publicly available on Kaggle \cite{ma2023eargateDataset}.

\textbf{ClearSpeech.} The dataset released as part of the ClearSpeech paper \cite{ma2024clearspeechDataset} supports speech enhancement research on earable platforms. The authors designed a custom earbud prototype with dual microphones—an in-ear microphone capturing occluded, bone-conducted speech and an out-ear microphone recording ambient air-conducted signals. Data were collected from 20 participants, each reading 300 sentences from the TIMIT corpus in quiet indoor conditions, yielding approximately 9.4 hours of clean dual-microphone recordings. To train denoising models, the authors also constructed a synthetic dataset by adding ambient noise from AudioSet and ESC-50 to clean recordings and simulating noisy in-ear signals using an out-to-in (O2I) transformation model. All data are synchronized and include paired clean and noisy audio, supporting supervised training of in-ear speech enhancement systems under realistic dual-microphone input. The dataset is publicly available on Kaggle \cite{ma2024clearspeechDataset}.


\textbf{Remarks.}
While this section reviews a limited number of publicly available datasets, broader adoption and reuse remain challenging. Unlike conventional data domains such as image or speech, earable datasets face unique difficulties due to their anatomical sensing context, modality diversity, and hardware variability. As a result, most datasets are tightly coupled to specific applications, limiting their generalizability across users, tasks, and systems. Below, we summarize four key limitations that currently hinder scalable and reproducible research:

\begin{itemize}
    \item \textbf{Contextual Fragility.}
    Earable signals are highly sensitive to user-specific and situational factors, including anatomical differences, wearing variation (e.g., insertion angle or seal tightness), motion state, and ambient noise. Even for the same user, small changes in posture, environment, or session can shift signal distributions, reducing intra-user consistency and cross-user generalizability. As a result, models trained on controlled datasets often underperform in real-world conditions, especially in applications like voice interaction or biometric authentication.
    \item \textbf{Cross-Dataset Compatibility.}
    While multimodal fusion (e.g., audio + IMU) and model pretraining are increasingly common, mismatches in sensor placement, sampling rate, format, and labeling schemes make it difficult to combine or reuse datasets. These incompatibilities raise concerns about whether modalities drawn from different datasets, such as microphones from one source and motion sensors from another, can be meaningfully aligned or integrated. This challenge often forces researchers to collect new data even for similar tasks.
    \item \textbf{Reproducibility Barriers.} 
    Many datasets are built on custom hardware, such as capacitive tips, piezo sensors, or bone-conduction microphones, which are rarely open-sourced or commercially available. Even with commodity components like MEMS microphones or IMUs, differences in placement, fit, or acoustic coupling may cause distribution shifts. These hardware inconsistencies, coupled with sparse documentation, missing calibration metadata, and unsynchronized signals, hinder reproducibility and fair cross-system evaluation.
    \item \textbf{Annotation and Task Granularity.}
    Some earable applications require complex, fine-grained annotations, such as jaw articulation, speech segments, or subtle behavioral events, that are difficult to label consistently across participants or modalities. Limited annotation standards, unclear task definitions, and sparse contextual metadata reduce label quality, making it harder to evaluate model generalization or compare across studies.
\end{itemize}

Together, these challenges highlight the need for better standardization in sensor setup, annotation protocol, and documentation practices. Addressing them will be essential to support broader reuse of datasets and accelerate progress in earable sensing research.


\subsection{Earable Signal Quality Assessment Tools}
\label{sec: Earable}

While datasets provide the foundation for algorithm development, the real-world reliability of earable sensing ultimately hinges on signal quality. In practice, signals captured by earables—such as in-ear audio, PPG, EEG, or IMU data—are highly susceptible to degradation from poor ear canal sealing, motion artifacts, and inconsistent sensor contact. However, many systems implicitly assume clean input, overlooking the dynamic and user-dependent variations in signal quality that occur during daily use. These degradations can significantly impair sensing accuracy and lead to unreliable or misleading downstream inferences in applications such as health monitoring, neural decoding, or attention tracking. As such, assessing and ensuring signal quality is not merely an optimization step, but a critical prerequisite for building robust and generalizable earable systems.

Demirel \etal ~\cite{demirel2024unobtrusive} addressed a frequently overlooked factor: the quality of ear canal sealing, which affects both in-ear microphone and photoplethysmography (PPG) signals. They proposed a novel, hardware-free method for estimating air leakage using the distortion patterns captured by in-ear microphones. By modeling acoustic distortion and applying machine learning, their system could detect poor fitting conditions in real time, offering a practical solution for unobtrusively monitoring ear fit and consequently improving sensing reliability. Jayas \etal ~\cite{jayas2024know} tackled the challenge of ensuring signal fidelity in ear-EEG systems, where body movements often disrupt electrode-skin contact. They proposed an in-situ channel selection method based on manifold learning, which extracts geometric features (dimension and curvature) from EEG signal manifolds to distinguish valid EEG channels from noise-contaminated ones. 
Complementing these efforts, Berent \etal ~\cite{berent2024quality} conducted a large-scale comparison of ear EEG with traditional scalp and intracranial EEG using clinical data. They demonstrated that ear EEG recorded during sleep exhibited higher signal quality than during wakefulness, with coherence and spectral similarities to scalp and intracranial signals, especially in low-frequency bands (delta, theta, alpha). Notably, contralateral ear EEG channels provided better signal resemblance than unilateral ones, confirming their relevance for high-quality neural monitoring.

\textbf{Remarks.} Signal quality assessment for earables is important for real-world deployment of ear-centric sensing systems. Future work should aim to develop unified signal quality metrics covering EEG, PPG, acoustic, and IMU signals simultaneously. Moreover, integrating signal quality assessment into closed-loop systems can enable dynamic reconfiguration of sensing parameters (\eg, gain, filtering, sensor fusion) in response to degradation. The inclusion of personalized signal quality baselines and cross-device generalizability remains an open challenge. Additionally, quality-aware machine learning models that explicitly model uncertainty due to signal degradation may improve robustness in health and cognitive inference tasks.

\begin{table}[t]
\caption{Summary of earable signal quality assessment.}
\label{tab: signal_quality}
\renewcommand{\arraystretch}{1.2}
\begin{tabular}{p{2.1cm}p{2cm}p{3.5cm}p{4.3cm}p{1.8cm}}
\hline
\textbf{System}  & \textbf{Sensor} & \textbf{Application} & \textbf{Advancement} &\textbf{Performance} \\ \hline
Demirel \etal  \cite{demirel2024unobtrusive} & In-ear mic & Ear fit monitoring & Hardware-free air leakage detection via acoustic distortion & Real-time poor fit detection \\ \hline
Jayas \etal  \cite{jayas2024know} & EEG & EEG signal quality assurance & In-situ channel selection via manifold learning & N/A \\ \hline
Berent \etal  \cite{berent2024quality} & EEG & Clinical signal comparison & Large-scale study on ear vs scalp/intracranial EEG & N/A \\ \hline
\end{tabular}
\end{table}
\section{Future Research Directions}
\label{sec: Future}

Around 2022, Choudhury~\cite{choudhury2021earable} and Röddiger \etal ~\cite{roddiger2022sensing} provided a visionary outlook on the future of earable computing from various dimensions, covering hardware platforms (e.g., sensor integration and on-device computation), power consumption (e.g., OS optimization and energy harvesting), multi-modal fusion (e.g., combining audio and inertial sensing), usability (e.g., form factor and weight for long-term monitoring), and the broader wearable-earable ecosystem (e.g., collaborative sensing with smartwatches). Their discussions have since inspired and guided much of the subsequent research in the field. For instance, platforms such as OmniBuds~\cite{montanari2024omnibuds} and OpenEarable 2.0~\cite{roddiger2025openearable} were developed as general-purpose earables that integrate multiple sensors and support on-device computation including deep learning inference. Additionally, the fusion of IMU and microphone sensors has enabled new capabilities (such as speech enhancement~\cite{he2023towards}) and enhanced performance in tasks like eating monitoring~\cite{wang2024gustosonicsense}. However, several challenges highlighted in the earlier discussions remain underexplored, including system-level optimizations for prolonged battery life and novel designs aimed at improving long-term usability.

We also conducted a chronological analysis of prior earable publications. From the application perspective, \textit{most existing research has focused on replicating functionalities} (such as gesture recognition, heart rate monitoring, and sleep tracking) \textit{that were previously demonstrated on smartphones or smartwatches}, often seeking improved accuracy or robustness. Only recently have we seen the emergence of applications uniquely suited to earables, such as ear disease monitoring and lung function assessment, which leverage their proximity to vital organs. Despite this progress, the full potential of earables as a human sensing platform remains largely untapped, and deeper exploration is warranted.

In response to the question raised in the introduction: \textit{earable computing is far from being a saturated research area. On the contrary, we believe it is entering an exciting new phase}. In the following sections, we outline five key research directions, \textbf{Hardware, Application, Software, Energy Consumption, and Usability} that call for further investigation in this rapidly evolving domain.

\subsection{Hardware}

While \Cref{subsec:hardware-platforms} has introduced a range of open-source earable platforms, practical deployment in real-world settings still poses significant hardware-related challenges. This section focuses on two core challenges: how to efficiently transmit high-fidelity sensor data, and how to support intelligent, low-latency processing through on-device AI inference.

\textbf{Wireless Data Transmission.}
Modern earables predominantly rely on Bluetooth connections, offering convenience by eliminating cables and enabling seamless smartphone integration. However, traditional Bluetooth protocols introduce several limitations for sensing applications. First, Bluetooth Classic typically supports only single-channel audio transmission at 16 kHz, which restricts applications such as ultrasound-based sensing \cite{duan2024f2key,zhang2023hpspeech} that require higher sampling rates (e.g., 48 kHz). Although the newer LE Audio \cite{LEAudio} standard improves capabilities, supporting dual-channel microphone streaming at 16 kHz per channel and potentially 48 kHz if fully implemented, it still depends on manufacturer support and remains insufficient for demanding applications like dual-channel ultrasound. Second, applications relying on multi-channel audio \cite{chatterjee2022clearbuds,ma2024clearspeech,han2024earspeech} or multimodal fusion \cite{duan2024f2key,choi2023earppg} often face bandwidth bottlenecks, particularly when continuous raw data streaming is needed. To address these challenges, one potential solution is to design hybrid transmission mechanisms, allowing the earable to dynamically switch between Bluetooth and wired modes depending on the application's bandwidth demands. High-bandwidth applications could leverage wired connections to ensure data fidelity and real-time throughput, while Bluetooth can be reserved for low-data-rate scenarios, prioritizing user mobility and comfort. Alternatively, future designs may offload buffered sensor data to the charging case during recharging, where higher-bandwidth links (like Wi-Fi) might be available for bundled transmission to the cloud.

\textbf{AI Inference Capability.}
Recent platforms like OpenEarable 2.0 \cite{roddiger2025openearable} have begun to integrate on-device AI accelerators to enable local inference, reducing reliance on external computation and improving real-time responsiveness. Specifically, OpenEarable incorporates a Tensilica HiFi 3z DSP Core dedicated for running neural network models efficiently at the edge. This architectural enhancement reflects a growing trend toward enabling more intelligent, autonomous processing within earables. However, despite these advances, the AI accelerators in current platforms are primarily optimized for lightweight models with constrained memory and computational budgets. For more complex tasks, such as high-fidelity audio enhancement or large-scale speech recognition, earables still face significant limitations, often necessitating model simplification or offloading to external devices. Continued progress in low-power AI hardware and system-level optimization remains crucial to fully unlock the potential of on-device intelligence for earable sensing.

\subsection{Application}
As sensing hardware and algorithms improve, earables are enabling new application domains that extend beyond the capabilities of traditional wearables. This section focuses on two key directions: applications uniquely enabled by the ear’s anatomy, and the shift from basic physiological metrics to richer, clinically meaningful biomarkers for long-term health monitoring.

\textbf{Earable-exclusive Applications.} 
While many earable applications have replicated functions from other wearables, such as heart rate monitoring or activity recognition, tasks uniquely suited to the ear remain underexplored. The ear’s anatomy, however, presents two promising sensing directions. First, its proximity to the brain and major blood vessels makes it an ideal site for neurophysiological sensing. Ear-EEG enables applications like sleep monitoring, mental load estimation, emotion recognition, and seizure detection. Yet, this comes with trade-offs: more electrodes improve spatial resolution but reduce comfort and increase design complexity, risking poor fit or signal instability \cite{kaongoen2023future}. Second, being near the respiratory tract and vocal apparatus, earables can capture breathing-related signals. They support applications like respiratory rate tracking, breathing mode classification, and lung function assessment, all without intrusive equipment like chest straps or nasal cannulas. In short, earables unlock access to physiological signals that are difficult to measure elsewhere, enabling a new class of exclusive sensing applications. Despite recent advancements, broader and novel applications remain worthy of exploration.

\textbf{Advanced Biomarker Monitoring.} 
Most existing earable applications focus on basic physiological metrics, such as heart rate for cardiovascular monitoring or breathing rate and mode for respiratory tracking. However, advancing toward deeper, clinically relevant biomarkers could transform earables from lifestyle accessories into credible health-monitoring tools. In cardiovascular assessment, for example, \cite{charlton2022assessing} has demonstrated that PPG can be used not only to track pulse rate but also to extract waveform-derived features such as rise time, pulse transit time, and pulse wave velocity. These indicators reflect arterial compliance and stiffness, key parameters in assessing vascular aging and identifying early signs of cardiovascular disease. In respiratory health, continuous monitoring of breath sounds using in-ear microphones may enable the detection of conditions such as asthma, chronic obstructive pulmonary disease, pneumonia, and sleep apnea. Acoustic features like wheezes, crackles, and apneic pauses serve as non-invasive markers of respiratory dysfunction. By facilitating passive, long-term data collection in daily life, earables could support early detection of respiratory disorders, reducing the need for frequent clinical visits and enabling more proactive health management.


\subsection{Software}
Beyond hardware, software plays a critical role in enabling robust and intelligent earable sensing. Modern systems increasingly depend on algorithms to fuse multimodal signals, coordinate across devices, and extract meaningful inferences. This section highlights key software directions, multi-sensor fusion, cross-wearable collaboration, and advanced modeling techniques that support scalable and adaptive earable applications.

\textbf{Multi-sensor Fusion.}
Integrating multiple sensing modalities within earable systems offers software-level opportunities to improve sensing accuracy, motion resilience, and physiological coverage. Through data fusion algorithms, complementary signals can be combined to perform cross-validation and artifact mitigation. For example, IMU data can help verify motion patterns and be used to suppress motion-induced noise in PPG or in-ear microphone signals via signal filtering or regression techniques. Beyond denoising, fusion pipelines can unlock new capabilities: recent work \cite{truong2022non} demonstrated that jointly analyzing PPG and in-ear microphone signals enables non-invasive blood pressure estimation by measuring vascular transit time, which leverages the different propagation speeds of acoustic and blood signals. Despite these benefits, fusion poses software challenges, including cross-modal synchronization, adaptive fusion strategies, and efficient processing frameworks. These must be carefully designed to operate within the resource constraints of earable systems.

\textbf{Cross-wearable Collaboration.}
Earables provide privileged access to both internal physiological signals and environmental context, but other wearable devices—such as smartwatches, smart glasses, and head-mounted displays—offer complementary sensing perspectives. While each platform has specific strengths, they also suffer from inherent limitations when used in isolation: smartwatches primarily capture arm and wrist motion but struggle with core posture or internal signals; smart glasses can detect eye and facial movements but are less effective for deep physiological monitoring; and head-mounted displays can track head pose or gaze but are bulky and energy-intensive. Cross-wearable collaboration offers a path to overcome these individual limitations by leveraging synchronized sensing across multiple devices. For instance, combining earables with smartwatches can enable more precise full-body activity or pose estimation; pairing with smart glasses allows for joint audio-gaze interaction and spatial awareness; integrating with AR headsets supports spatial audio and immersive context-aware communication. Enabling such cooperative sensing across wearables will require robust time synchronization, efficient data fusion protocols, and energy-aware communication strategies designed for heterogeneous, resource-constrained platforms.

\textbf{Advanced Data Modeling Techniques.}
Precise sensing in earable applications relies not only on hardware but also heavily on effective modeling techniques. Early works primarily used classical signal processing methods and traditional machine learning models tailored to specific tasks. More recently, self-supervised learning and transfer learning have enabled better feature extraction from unlabeled data and cross-modal adaptation, improving robustness under real-world conditions. A growing trend is the adoption of foundation models for physiological sensing. Instead of training separate models for each application, researchers increasingly use large pretrained models that can be adapted or fine-tuned for new tasks with limited data. For example, PaPaGei \cite{pillai2024papagei} introduces a foundation model for PPG, pretrained on over 57,000 hours of data, and shows strong performance across various health tasks. These developments suggest that modular, reusable models and pretraining strategies could significantly accelerate progress in earable sensing. Future work should explore multi-modal foundation models, domain-adaptive pretraining (across audio, IMU, PPG), and lightweight model architectures optimized for on-device inference.

\subsection{Energy Consumption}

Power efficiency remains a fundamental constraint for earable devices due to their limited size and battery capacity. This section outlines how sensor selection, computation offloading strategies, and power asymmetry across earbuds influence energy consumption, and discusses possible design strategies to extend battery life while preserving functionality.

\textbf{Sensor and Algorithm Selection.} 
Many earable applications can be achieved with different sensing principles and modalities. For example, in facial gesture recognition, IMU-based approaches \cite{zhao2024ui} generally consume less power but may offer lower accuracy compared to microphone-based \cite{yang2024maf} solutions. However, each sensor type comes with its sampling rate, algorithmic complexity, and hardware demands. Some applications require sensor arrays or multimodal fusion, all of which can significantly increase energy consumption. In academic research, even for the same application, optimization goals often vary—some prioritize accuracy, others focus on robustness to noise, and still others aim to minimize power or latency. In real-world scenarios, however, developers must carefully navigate these trade-offs. When a slight reduction in accuracy is acceptable, it may be more practical to choose a lower-power modality or a simplified algorithm to extend battery life and improve long-term usability. Therefore, a thorough investigation and benchmarking across different sensing modalities are necessary before a given application can be efficiently deployed in real-world settings.

\textbf{Computation Offloading Strategy.} 
Due to the inherent limitations of earables, including restricted memory, limited processing power, and small battery capacity, many practical systems offload computation to smartphones or cloud servers. While this enables more complex models and reduces on-device load, it introduces several challenges. First, continuous wireless transmission (especially of audio or high-frequency signals) can significantly increase power consumption. Second, offloading introduces latency, which may hinder real-time applications. Third, and most importantly, transmitting raw physiological data, such as breathing sounds, EEG, or in-ear audio, raises serious privacy concerns, as such data can reveal sensitive health or behavioral information. Although systems often adopt empirical rules for deciding when to offload, no standard framework exists. Recent efforts have explored lightweight on-device pre-processing, such as feature extraction or anonymization, to reduce both data volume and privacy risk. Building adaptive strategies that balance performance, efficiency, and privacy based on context remains an open challenge for future work.

\textbf{Battery Balance.}
As previously discussed, most Bluetooth earphones use an asymmetric architecture where one side, typically the primary earbud, handles microphone input and communicates with the smartphone, while the secondary connects via a low-power internal link such as BLE \cite{Alexander2025Smart}. This simplifies communication but often causes faster battery drain on the primary side. A similar imbalance occurs in on-device computation, where tasks may be assigned to one earbud or unevenly distributed based on sensor location or system design. While sufficient for simple applications, this becomes problematic for binaural tasks like beamforming, spatial localization, or EEG asymmetry analysis, where early depletion of one side can impair function. To address this, some systems \cite{hurwitz2020cell} explore dynamic role-switching and workload balancing based on battery status or sensing needs, or support task migration to the better-powered side. Still, maintaining battery symmetry remains an open challenge, especially in long-duration, high-demand scenarios.

\subsection{Usability}
Beyond technical performance, the success of earables depends heavily on how well they integrate into users’ daily lives. This section evaluates usability considerations including long-term hearing aid usage, the influence of audio playback features such as ANC and transparency, and the design trade-offs of open-ear versus in-ear form factors, all of which impact comfort, compliance, and sensing robustness.

\textbf{Hearing Aid Usability.} 
Earables encompass a range of devices, including earbuds, hearing aids, and bone-conduction headphones. Yet most research has focused on earbuds, with hearing aids remaining relatively underexplored in the sensing community. Despite this, hearing aids offer unique advantages that make them highly promising for long-term physiological and behavioral monitoring. Unlike earbuds, hearing aids are typically used throughout the day \cite{wearHearingAids}, making them well-suited for continuous sensing. This persistent wear supports fine-grained tracking of vital signals, activity patterns, and long-term health trends. Many users are elderly or hearing-impaired, and could particularly benefit from context-aware applications such as fall detection, cognitive monitoring, hearing environment adaptation, or social engagement assessment. Hearing aids may also enable real-time speech clarification, hearing training, and voice-based reminders to assist with daily living. However, hearing aids are often highly personalized, customized to each user’s audiometric profile, ear canal shape, and comfort preferences. This process, involving hearing threshold fitting and acoustic sealing \cite{alamdari2020personalization}, can hinder the deployment of general-purpose sensing systems. Earable applications for hearing aids should therefore be designed to reduce enrollment burden and support generalizability across diverse user populations.

\textbf{Impact of ANC and Transparency.}
Active Noise Cancellation (ANC) and Transparency Mode have become standard features in modern earables, allowing users to adapt their auditory experience across different environments. However, these audio processing modes can significantly interfere with sensing applications that rely on playback signals and recorded echoes, such as those used for speech enhancement, authentication, or respiratory monitoring. For instance, ANC may introduce frequency-dependent colorations or phase shifts, while Transparency Mode may alter echo profiles due to signal mixing from internal and external microphones. These modifications can distort the original audio cues or feedback patterns critical for sensing accuracy. Despite their prevalence, it does not remain easy to characterize or compensate for such effects in practice. This is due in part to the proprietary nature of ANC/transparency implementations, which are not publicly documented, and also to the use of varying algorithms across device models, often dictated by microphone placement and vendor-specific designs. As a result, applications involving audio feedback or echo analysis must explicitly account for these transformations, though doing so remains an open and underexplored challenge.

\textbf{Open-ear Form Factor.}
Most early earable devices adopt an in-ear, occlusive form factor. However, prolonged ear canal occlusion may pose health risks \cite{mazlan2002ear}, including increased humidity, microbial growth, and earwax buildup. In response, open-ear designs, such as bone-conduction \cite{openrunpro2} and ear-clip \cite{BoseUltraOpenEarbuds} earphones have emerged, offering improved breathability and long-term comfort. Yet, this shift introduces new challenges. Many physiological and activity-sensing applications leverage the occlusion effect to enhance low-frequency body-conducted sounds. Open-ear designs, by leaving the ear canal open, lose this acoustic amplification advantage. Additionally, the altered positioning and looser fit of open-ear devices complicate sensor deployment: for instance, PPG sensors may lose skin contact, and IMUs on ear-clip headphones may be more susceptible to motion artifacts due to increased movement. Due to these limitations, only a small number of existing studies \cite{he2024hcr} have specifically targeted sensing with open-ear earphones. This highlights the significant design and engineering barriers that must be addressed before such devices can support a wider range of physiological or behavioral sensing applications.

\section{Conclusion}
\label{sec:conclusion}
This survey presents a comprehensive update on the state of earable computing from 2022 to 2025. By analyzing over one hundred recent publications, we systematically compare new findings with prior efforts, highlighting key advancements in sensing principles, system performance, hardware integration, and deployment feasibility. Despite these promising developments, earable computing still faces critical challenges around hardware usability, energy efficiency, and real-world generalizability, which demand thoughtful system design and user-centric engineering to ensure seamless daily use. Looking ahead, we believe earables are poised to become a cornerstone of next-generation ubiquitous computing. With their unique ability to passively and continuously capture both internal physiological signals and external contextual cues, earables represent a rich sensing platform for intelligent, human-centered applications. By addressing the remaining gaps and exploring underutilized opportunities, future research can further elevate earables from peripheral accessories to essential components of health monitoring, interaction, and ambient intelligence systems.

\bibliographystyle{ACM-Reference-Format}
\bibliography{references}

\end{document}